\documentstyle[aasms4,epsfig]{article}  

\newcommand{\be}{\begin{equation}}
\newcommand{\ee}{\end{equation}}
\newcommand{\bea}{\begin{eqnarray}}
\newcommand{\eea}{\end{eqnarray}}
\newcommand{\nn}{\nonumber}

\newcommand{\noi}{\noindent}
\newcommand{\appgeq}{\stackrel{>}{\sim}}
\newcommand{\appleq}{\stackrel{<}{\sim}}

\newcommand{\Bphi}{B_\phi}
\newcommand{\Bz}{B_z}
\newcommand{\Br}{B_r}

\newcommand{\Ps}{P_S}
\newcommand{\Pave}{\langle P\rangle}
\newcommand{\pave}{\langle p \rangle}
\newcommand{\aave}{\langle a_4/a \rangle}
\newcommand{\Rs}{{R_S}}
\newcommand{\Mbe}{M_{B.E.}}
\newcommand{\Pbe}{P_{B.E.}}

\newcommand{\Gz}{\Gamma_z}
\newcommand{\Gphi}{\Gamma_\phi}

\newcommand{\bp}{b_\phi}
\newcommand{\Aphi}{A_\phi}
\newcommand{\Phicl}{\Phi_{cl}}

\newcommand{\cross}{{\bf\times}}

\lefthead{Fiege \& Pudritz}
\righthead{Prolate Cores in Filamentary Molecular Clouds}

\begin{document}

\title{Prolate Cores in Filamentary Molecular Clouds}
\author{Jason D. Fiege and Ralph E. Pudritz}
\affil{Dept. of Physics and Astronomy, McMaster University, \\
1280 Main St. W., Hamilton, Ontario, L8S 4M1\\
email: fiege@physics.mcmaster.ca, pudritz@physics.mcmaster.ca}

\begin{abstract}
We present a new model of molecular cloud cores that originate from filamentary clouds 
that are threaded by helical magnetic fields.  Only modest toroidal fields are required to produce
elongated cores, with intrinsic axis ratios in the range $0.1\appleq q\appleq 0.9$ and mean projected axis ratios in
the range $0.3\appleq \pave \appleq 1$.  Thus many of our models are in good agreement with the observed shapes
of cores (Myers et al 1991, Ryden 1996), which find axis ratios distributed about the mean values
$\pave\approx 0.5-0.6$ and $q\approx 0.4-0.5$.  
We find that the Bonnor-Ebert critical mass is reduced by about 20\% by the helical field threading our models.
We also find that models are generally most elongated when the mass is significantly less than the Bonnor-Ebert critical 
mass for unmagnetized cores.
\end{abstract}


\section{Introduction}
\label{sec:intro}

There is considerable observational evidence that most molecular cloud cores and Bok globules are elongated structures that may
be approximately prolate in shape.  Myers et al. (1991) measured the projected axis ratios of 16 cores in line
intensity maps of $NH_3$, $CS$, and $C^{18}O$ and found that the mean projected axis ratio $\pave$,
defined as the ratio of the minor to major axis, is in the range $0.5-0.6$.  They assumed
that the cores are randomly oriented on the sky and calculated the allowed values of the intrinsic
axis ratio $q$.  Their analysis showed that modest elongation with $q=0.4-0.5$ can account for the projected axis ratios
if the cores are intrinsically prolate, but rather extreme flattening, with $q<0.1-0.3$ is required if they are
intrinsically oblate.  Moreover, 6 of the 16 cores appear to be inside of and aligned with larger filamentary
structures of lower density observed in CO or dust extinction.  They argued that cores must almost certainly be prolate
in such cases, which are actually quite common.
The Schneider and Elmegreen (1979) Catalogue of Dark Globular Filaments shows many other
examples of elongated cores contained within filamentary structures.

Ryden (1996) extended the pioneering work of Myers et al. (1991) by analyzing the distribution of axis ratios for a
large sample of Bok globules and cloud cores taken from several independent data sets (See Ryden for references).  She
found that randomly oriented oblate cores are inconsistent with the observed axis ratios for most samples to a high
confidence level ($>99$\% in some cases).  On the other hand, she found that prolate cores are almost always
consistent with the data.
Her analysis predicts that the mean intrinsic axis ratio for cores is typically 0.4-0.5, in agreement with Myers et al. (1991), but also that
there should be significant numbers of cores that are either significantly more elongated or more nearly spherical.

If cores are truly prolate, then several interesting questions are raised about the non-isotropic forces responsible
for their shapes.  It is well established that rotation is negligible compared to magnetic and gravitational forces (Goodman et al. 1993),
and tends to make cores oblate in any case (Tomisaka, Ikeuchi, and Nakamura 1988a, hereafter TIN).  
It is known, however, that large scale magnetic fields play a central
role (Myers and Goodman 1988a,b) in supporting cores against self-gravity.  Considerable theoretical
effort has gone into producing magnetized, non-spherical models of cloud cores (Mouschovias 1976a,b, TIN,
Tomisaka, Ikeuchi, and Nakamura 1988b, 1989, 1990 (hereafter TIN88b, TIN89, and TIN90), and Tomisaka 1991).
Most of these models
assume that the magnetic field threading the core is purely poloidal.
However, purely poloidal fields result in oblate
structures which, as we have seen, do not seem to explain the observations.

We show that prolate models of cores, whose shapes are in better agreement with the observations,
are readily produced by including a toroidal component of the
magnetic field, so that the field is helical in general
\footnote{
We use the term ``helical'' rather loosely to describe any field geometry that includes both poloidal and toroidal field components.
The field lines in our prolate core models are not true helices in the mathematically correct sense.
}. 
The basic point is that
the radial pinch of the toroidal field helps to squeeze
cores radially into a prolate shape, while helping to support the gas along the axis of symmetry.
Tomisaka (1991) finds that toroidal fields can result in moderately prolate
equilibria with intrinsic axis ratios of $\appgeq 2/3$.  Therefore, his most prolate models are
less elongated than the mean axis
ratio predicted by the observations.  His prolate equilibria are also only obtained
when the cores are not very centrally concentrated, with a density contrast of only $\sim 3$ between the
centre and the surface.  We shall show that our models are both more elongated and more centrally concentrated.

One way that a core could acquire a helical field is by inheriting it from
its parent filamentary cloud.
We previously derived models of filamentary clouds threaded by helical magnetic fields
in Fiege and Pudritz 1999a, (hereafter FP1).
We showed that helical fields with constant poloidal and toroidal flux to mass loading are consistent
with both the virial properties of filaments and the existing data on their density profiles (Alves et al.
1998, Lada et al. 1998).
Our main goal in this paper is to determine the equilibrium states that are accessible to cores
that might form from elongated fragments of these magnetized filaments.
Three parameters are required to specify our filament models, which we discuss in the next section.
In addition to these three parameters, we only need to specify the mass of a core to calculate its equilibrium structure
in our model.

Our main findings are that a large range of core models with projected axis ratios $\pave$
ranging from $\sim 0.3$ to $\sim 1$ can be obtained
from our models of filamentary clouds threaded by helical fields.
Our converged solutions define the prolate magnetic analogues of Bonnor-Ebert (1956, 1955) spheres.
Their magnetic structure is characterized by a relatively strong ``backbone''   
of poloidal flux surrounded by a region where the toroidal field dominates,
but usually does not exceed the peak poloidal field.

We rather completely map out our parameter space by searching for models whose parent filaments obey the
observational constraints discussed in FP1, and whose fragmentation is driven predominantly by self-gravity
rather than by MHD sausage instabilities (See Fiege \& Pudritz 1999b for details, hereafter FP2).
The prolate core models that converge are found to occupy
a well-defined region of parameter space.  The failure of models
to converge outside of this region can be traced to non-equilibrium resulting
from gravitational and MHD instabilities.
We also show that the magnetic analogue of the Bonnor-Ebert critical mass
is somewhat reduced by the helical field in our models.

A brief outline of our paper is as follows.  In Section \ref{sec:SCFM} we outline the self-consistent field method that we use to
generate our equilibria.  We describe our numerical method in Section \ref{sec:numerical},
and show a gallery of representative models in Section \ref{sec:MC}.
In Section \ref{sec:sequences}, we show sequences
of models with varying mass and external pressure, and show that the Bonnor-Ebert critical mass is reduced
by the magnetic field.
We derive a singular model of isothermal cores wrapped by purely toroidal
fields in Section \ref{sec:SIE}, and discuss our results in Section \ref{sec:discussion}.

\section{The Self-Consistent Field Method}
\label{sec:SCFM}

Our basic strategy is to use a relatively minor generalization of the self-consistent field method
originally developed by Mouschovias (1976a), (subsequently used by
Mouschovias 1976b, TIN, TIN88b, TIN89, TIN90, and Tomisaka 1991) to follow the relaxation
of finite fragments of the filamentary clouds discussed in FP1.  The reader is referred to
TIN for a full derivation of the equations.  We briefly outline the method
and discuss our modifications in this section.

The self-consistent field method iteratively solves Poisson's equation and the equations of magnetohydrostatic equilibrium
for a self-gravitating cloud threaded by a magnetic field and truncated by an external pressure.  The source of the
external pressure, which is presumably relatively low density molecular or atomic gas, is assumed to be
non-self-gravitating and of negligible density.  Thus, the external pressure $\Ps$ serves mainly to
define the surface of the cloud in our model.

The self-consistent field method separates the equations of magnetohydrostatic equilibrium into equations describing the
detailed balance of gravitational, magnetic, and hydrostatic forces parallel to the poloidal field lines and perpendicular to them.
Following TIN, we use the flux quantity $\Phi$ which is proportional to the true magnetic flux $\Phi_{mag}$:
\be
\Phi_{mag}=2\pi\Phi.
\label{eq:Phimag}
\ee
The equation of magnetohydrostatic equilibrium is given by
\be
\nabla P + \frac{{\bf B}\cross\left(\nabla\cross{\bf B}\right)}{4\pi} + \rho\nabla\Psi = 0.
\label{eq:magnetostatic}
\ee
Writing the toroidal field as
\be
\Bphi=\frac{\bp}{r},
\label{eq:bphi}
\ee
it follows directly from the azimuthal component of equation \ref{eq:magnetostatic} that
$\bp$ is constant along field lines (See TIN for proof); hence $\bp$ is a function of $\Phi$ alone.

The magnetic field exerts no force in the direction parallel to the poloidal field lines.  Thus the
equation of magnetohydrostatic equilibrium reduces to the hydrostatic equation along a field line:
\be
\frac{\partial P}{\partial s}+\rho \frac{\partial \Psi}{\partial s}=0,
\label{eq:hydro}
\ee
where $s$ is the distance along the field line ($s=0$ at the symmetry plane defined by $z=0$), $P$ is the
effective pressure, which may contain a contribution due to non-thermal motions of the gas, $\rho$ is the density, and
$\Psi$ is the gravitational potential (We do not include the effects of rotation in our analysis.).
We assume that the gas is
``isothermal'' in the sense that the total velocity dispersion $\sigma$ is constant:
\be
P=\sigma^2\rho.
\label{eq:EOS}
\ee
In this case, equation \ref{eq:hydro} can be integrated to obtain
\be
\rho=\frac{Q(\Phi)}{\sigma^2} e^{-\Psi/\sigma^2},
\label{eq:hydro2}
\ee
where $Q(\Phi)$ is constant along any field line
\footnote{
The field constant $Q$ is related to the Bernoulli integral along a field line, which is just $\sigma^2 \ln Q$.
}, 
but may vary across them.
We note that our $Q$ is identical to the $q$ used by TIN; we use this different notation
to avoid confusing $Q$ with the intrinsic axis ratio $q$ of a core.
The poloidal components of the magnetic field are related to the flux $\Phi$ by
\bea
B_r=-\frac{1}{r}\frac{\partial\Phi}{\partial r} \nn\\
B_z=\frac{1}{r}\frac{\partial\Phi}{\partial z}.
\label{eq:bpol}
\eea

With the help of equations \ref{eq:bphi} and \ref{eq:hydro2}, the magnetostatic equilibrium equation
(equation \ref{eq:magnetostatic}) can be re-arranged to give an elliptic type partial differential 
equation for the azimuthal component of the magnetic vector potential $\Aphi$ (see TIN, equation 2.26):
\be
\Delta_1(r\Aphi)=-4\pi r^2 \frac{dQ}{d\Phi} e^{-\Psi/\sigma^2}-\frac{1}{2}\frac{d\bp^2}{d\Phi},
\label{eq:fund}
\ee
where $\Aphi$ is related to the flux by
\be
\Aphi=\frac{\Phi}{r},
\label{eq:PhiAphi}
\ee
and the differential operator $\Delta_1$ is given by
\be
\Delta_1=\frac{\partial^2}{\partial z^2}+r\frac{\partial}{\partial r}\left(\frac{1}{r}\frac{\partial}{\partial r}~~\right).
\label{eq:delta1}
\ee
We write the left hand side of equation \ref{eq:fund} in terms of $\Aphi$, rather than $\Phi$, since this is the form
of equation \ref{eq:fund} that we actually finite difference in Section \ref{sec:numerical} and
solve in Sections \ref{sec:MC} and \ref{sec:sequences}.

Finally, we require Poisson's equation for the gravitational potential:
\be
\nabla^2 \Psi=4\pi G \rho,
\label{eq:poisson}
\ee
where $\rho$ is related to $Q$ and $\Psi$ by equation \ref{eq:hydro2}.

It remains only to specify the functions $Q(\Phi)$ and $\bp(\Phi)$ to close the system of equations given above.
TIN derive the functional form of $Q(\Phi)$ by assuming that cores relax to equilibrium from initially
uniform, spherical clouds threaded by a constant poloidal magnetic field.
We assume instead that cores relax from
ellipsoidal fragments of a parent filamentary cloud that is threaded by a helical magnetic field
(See FP1).
Therefore, we extend the work of Mouschovias (1976a,b) and TIN by examining the equilibrium states of
cores that relax from filamentary, rather than spherical, initial clouds.

The mass contained between poloidal flux surfaces defined by $\Phi$ and $\Phi+\delta\Phi$ is given by
\be
\delta M=\int_0^{s_m} 4\pi r\rho \delta w ~ds,
\label{eq:q1}
\ee
where $s$ is the distance along the field line, and $s_m$ is its maximum value at the
surface of the core, where the pressure drops to $\Ps$.  The thickness $\delta w$ of the flux tube varies
along its length in such a way that the flux $\delta\Phi$ between the flux surfaces remains constant:
\be
\delta w = \frac{\delta\Phi}{r |B_{pol}|}
\ee
where $|B_{pol}|=\sqrt{\Br^2+\Bz^2}$ is the magnitude of the poloidal magnetic field.
Combining this equation with equations \ref{eq:hydro2} and \ref{eq:q1}, we easily obtain the mass to flux ratio:
\be
\frac{dM}{d\Phi}=\frac{4\pi Q(\Phi)}{\sigma^2}\int_0^{s_m}\frac{e^{-\Psi/\sigma^2}}{|B_{pol}|} ds,
\ee
where we have used the fact that $Q$ is constant along field lines to take it outside the integral.
Solving for $Q(\Phi)$ and replacing the integration over $s$ with an equivalent integral over $z$,
we obtain
\be
Q(\Phi)=\frac{\sigma^2 dM/d\Phi}{4\pi}\left[ \int_0^{z_m} \frac{e^{-\Psi/\sigma^2}}{\Bz} dz \right]^{-1},
\label{eq:Q}
\ee
where the integral is to be performed along a poloidal field line, and $z=z_m$ at the surface of the cloud.

In order to use equation \ref{eq:Q}, we need to evaluate the mass to flux ratio $dM/d\Phi$
as a function of $\Phi$ for the initial filament, since this function is preserved by flux-freezing as the core
settles toward its equilibrium state.
As discussed in FP1, our models of magnetized filamentary molecular clouds require 3 parameters, namely the
poloidal and toroidal flux to mass ratios $\Gz$ and $\Gphi$ and a third parameter $C$ that determines the radial
concentration of the cloud.	The poloidal and toroidal magnetic fields are given by our filamentary cloud model as
\bea
\Bz &=& \Gz\rho \nn\\
\Bphi &=& \Gphi r \rho.
\label{eq:fields}
\eea
The surface of the filament, where the
pressure drops to that of the ISM, is located at radius \be
\Rs=r_0 e^C.
\ee
where $r_0$ is the core radius defined by
\be
r_0=\frac{\sigma^2}{4\pi G\rho_c}
\label{eq:r0}
\ee
and $\rho_c$ is the central density of the filament.
We consider an ellipsoidal ``fluctuation'' centred on the axis of the filament and assume that all of the gas initially inside of
the ellipsoid becomes part of the core in its equilibrium state.
We assume that the semiminor axis of the ellipsoid is equal to the radius $\Rs$ of the filament. The semimajor axis $Z_0$
(in the direction of the filament axis) is then determined by the mass chosen for the core, as we now demonstrate.

The poloidal field lines threading a filament run parallel to the filament axis since there is no radial magnetic
field.  Therefore, surfaces of constant flux are concentric cylinders centred on the filament axis.  The mass and flux contained
within the ellipsoidal fluctuation between cylindrical radii $r$ and $r+\delta r$ are given respectively by
\bea
\delta M &=& 4\pi r\rho Z_0\sqrt{1-r^2/\Rs^2}\delta r \label{eq:deltaM} \\
\delta\Phi &=& \Gz r\rho ~\delta r.  \label{eq:deltaPhi}
\eea
The second of these equations can be integrated to give the total flux $\Phi$ threading the ellipsoid.
We find that $\Phi$ is related to the mass per unit length $m$ of the filament by
\be
\Phicl=\frac{\Gz m}{2\pi}.
\ee
The mass to flux ratio is obtained by combining equations \ref{eq:deltaM} and \ref{eq:deltaPhi}:
\be
\frac{dM}{d\Phi}=\frac{4\pi Z_0}{\Gz} \sqrt{1-r^2/\Rs^2}.
\label{eq:dmdphi}
\ee
We note that this equation gives $dM/d\Phi$ as a function of $\Phi$ alone (numerically), 
as required by equation \ref{eq:Q}, if we consider $r$ to be a function of $\Phi$ within the filament.
The semimajor axis $Z_0$ which appears in equation \ref{eq:dmdphi} is obtained in terms of the mass and flux by integrating
this equation and rearranging:
\be
Z_0=\frac{\Gz M}{4\pi} \left(\int_0^{\Phicl}\sqrt{1-r^2/\Rs^2} ~d\Phi \right),
\label{eq:Z0}
\ee
which is calculated numerically.

Amp\`ere's law shows that the function $\bp(\Phi)$ is directly proportional
to the total poloidal current passing through the flux surface threaded by flux $\Phi$.
Since $\bp$ is a function of $\Phi$ alone,
the poloidal currents associated with the toroidal field must flow only in the direction of the poloidal field.
Tomisaka (1991) assumed that $\bp\propto\Phi$.
However, we obtain the function $\bp(\Phi)$ in our model by assuming that cores form without dissipation from
the initial filamentary state, so that the current flowing between any two flux surfaces is a conserved quantity.
Specifically, the functional form of $\bp(\Phi)$ follows from equations \ref{eq:bphi} and \ref{eq:fields}:
\be
\bp(\Phi)=\Gphi r^2\rho,
\label{eq:b}
\ee
where $r$ and $\rho$ are the radius and density of the parent filament, considered to be functions of the flux $\Phi$.

\subsection{Dimensionless Equations}
\label{sec:dimensionless}
Following FP1, we write our equations in dimensionless form by normalizing all quantities with respect to the
velocity dispersion $\sigma_{fil}$ of the filament and the density $\rho_c$ at its centre.  The dimensional scale
factors can then be written as follows:
\bea
\sigma_0^2 &=& \sigma_{fil} \nn\\
\rho_0 &=& \rho_c \nn\\
r_0^2 &=& \frac{\sigma_{fil}^2}{4\pi G\rho_c} \nn\\
\Psi_0 &=& \sigma_{fil}^2 \nn\\
P_0 &=& Q_0 = \sigma_{fil}^2\rho_c \nn\\
B_0 &=& P_0^{1/2} \nn\\
A_0 &=& b_{\phi,0} = r_0 B_0 \nn\\
\Phi_0 &=& r_0^2 B_0.
\label{eq:dimensionless}
\eea
The dimensionless forms of the flux to mass ratios $\Gz$ and $\Gphi$ are given in FP1 (equation 33) as
\bea
{\tilde\Gz} &=& \sqrt{\frac{\rho_c}{\sigma^2}}\left(\frac{\Bz}{\rho}\right) \nn\\
{\tilde\Gphi} &=& \frac{1}{\sqrt{4\pi G}}\left(\frac{\Bphi}{r\rho}\right).
\eea

Normalizing all quantities with respect to these dimensional scaling factors,
only Poisson's equations changes form:
\be
\nabla^2 \Psi=\rho.
\label{eq:poisson2}
\ee
The dimensionless forms of the remaining equations are identical to their dimensional forms given
by equations \ref{eq:bphi}, \ref{eq:EOS}, \ref{eq:hydro2}, \ref{eq:bpol}, \ref{eq:fund}, \ref{eq:PhiAphi},
and \ref{eq:Q}, where (dimensionless) $\sigma=1$ is specified as an input parameter
\footnote{
We only consider the case where $\sigma$ is the same in the core as in the parent filament (dimensionless
$\sigma=1$).  However, we leave $\sigma$ in our equations so that this condition might be relaxed in future work.
}
, and
$\rho$, $P$, $\Psi$, $\Aphi$, $\Phi$, ${\bf B}$, and $Q$ are all solved self-consistently.

\subsection{Boundary Conditions}
We assume that the poloidal magnetic field threading the original filament is
continuous at its surface and constant in the external medium, so that $\Aphi\propto r$ in the external medium.
The total flux threading our grid, of radius $r_{max} \gg \Rs$, is given by
\be
\Phi_{max}=\frac{\Gz}{2\pi}\left[m+\pi (r_{max}^2-\Rs^2)\Ps/\sigma^2\right].
\ee
Therefore, $A_{\phi,max}$ is given by
\be
A_{\phi,max}=c r_{max},
\ee
where $c$ is a constant:
\be
c=\frac{Gz}{2\pi r_{max}^2} \left[m+\pi (r_{max}^2-\Rs^2)\Ps/\sigma^2\right].
\ee
We find it convenient, for computational reasons, to define a ``modified'' vector potential $A$ such that
\be
A=\Aphi-cr.
\label{eq:AAphi}
\ee
It is easily verified that $A$ satisfies the same equation (equation \ref{eq:fund}) as $\Aphi$.
However, $A$ vanishes at the outer radius of our grid, where $r=r_{max}$.

To obtain the boundary condition on $A$ at large $z$, we assume only that there is no radial component 
of the magnetic field at the edge of our grid corresponding to $z=z_{max}$
\footnote{
This differs slightly from TIN, who assume that the magnetic field connects to their initially uniform field at large $z$.
}.
The boundary conditions on $\Psi$ and $A$ (in our dimensionless variables) are as follows.  \\
$r=0:$
\begin{equation}
\partial\Psi/\partial r=0, ~~~ A=0
\end{equation}
$r=r_{max}:$
\begin{equation}
\Psi=\frac{M}{4\pi\sqrt{r_{max}^2+z^2}}, ~~~ A=0
\end{equation}
$z=0:$
\begin{equation}
\partial\Psi/\partial z=0, ~~~ \partial A/\partial z=0
\end{equation}
$z=z_{max}:$
\begin{equation}
\Psi=\frac{M}{4\pi\sqrt{r^2+z_{max}^2}}, ~~~ \partial A/\partial z=0
\end{equation}

\section{Numerical Method}
\label{sec:numerical}
We briefly describe our numerical implementation of the self-consistent field method in this section.
We work in modified radial coordinates defined by
\bea
r &=& S_r \tan{\xi} \nn\\
z &=& S_z \tan{\eta}
\eea
so that radial infinity and infinity along the axis of symmetry are respectively mapped to $\xi=\pi/2$ and $\eta=\pi/2$.
Our grid is uniformly spaced in $\xi$ and $\eta$ and contains $2^7+1=129$ points in each dimension.  The advantage of working in these modified
coordinates is that they provide a good dynamical range (spatially) for our calculation.  
We obtain good resolution within the core by keeping a large fraction
(typically $\sim 1/3-1/2$) of the grid cells within its boundary, but our grid still extends to large radius where the boundary conditions stated in the
previous section apply.

We begin by initializing our numerical method as follows.
We find an equilibrium solution for a magnetized filament using the method discussed in FP1.
From this solution, we calculate $dM/d\Phi$, $Z_0$, and $\bp(\Phi)$ using equations 
\ref{eq:dmdphi}, \ref{eq:Z0}, and \ref{eq:b} respectively.
We begin with a initial guess of the density distribution $\rho(r,z)$ (eg, truncated Gaussian), as well
as the magnetic vector potential $A(r,z)$, which gives $\Phi(r,z)$ by
equations \ref{eq:PhiAphi} and \ref{eq:AAphi}.
We then solve our system of equations iteratively as follows:\\
1) We finite difference equation \ref{eq:poisson2}, and solve for the gravitational potential $\Psi(r,z)$\\ 
2) Next, we calculate the poloidal magnetic field using equation \ref{eq:bpol}, and accurately locate the
surface of the core by finding the surface where the pressure drops to $\Ps$.\\
3) We integrate along poloidal field lines (level surfaces of $\Phi$) to solve equation \ref{eq:Q} for $Q(\Phi)$.  We differentiate $Q(\Phi)$ to
obtain $dQ/d\Phi$, which is one of the source terms in equation \ref{eq:fund}. \\
4) We interpolate the functions $Q(\Phi)$, $dQ/d\Phi$, and $d\bp^2/d\Phi^2$ over our grid, using the current value of $\Phi$ at each grid point.\\
5) We update $A(r,z)$ by solving the finite difference equation obtained from equation \ref{eq:fund}, using the 
functions calculated in steps 1 and 5.

We repeat steps 1-5 until the solution either converges or is rejected.
We keep track of the maximum relative change in density for all cells whose density is greater than half of the central density.
We consider a solution to have converged when this quantity either becomes less
than $1\times 10^{-5}$, or it remains below $1\times 10^{-4}$ for
50 consecutive iterations.  A solution is rejected if the central density falls
below that of the original filament ($\rho=1$ in our dimensionless units), or appears to be growing without bound.
Our code allows a maximum of 750 iterations; this limit is never reached in practice
since solutions that converge usually do so within $\sim 250$ iterations.

\section{Exploration of the Parameter Space}
\label{sec:MC}
In this section we explore the range of equilibrium models that are accessible to finite segments of the filament
models described in FP1.  We have selected those models from FP1 that agree with the observational constraints obtained
from our virial analysis of the available observational data (see FP1 equation 22), and whose kinetic and magnetic energies are nearly
in equipartition, with $0.5 \le M/K \le 2$.  Moreover, we have selected only filaments that fragment slowly, with growth
timescales
\be
\tau_{frag} \ge 1.8 \left(\frac{n_c}{10^4~cm^{-3}}\right)^{-1/2}~Myr
\label{eq:tauf}
\ee
We refer the reader to FP2 for a discussion of fragmentation in filamentary clouds
and the normalization used in equation \ref{eq:tauf}.
The reason for this selection is to exclude filament models that are dominated
by rapidly growing MHD-driven sausage instabilities, rather than gravitational instabilities.
A total of 71 filament models from FP1 and FP2 meet these criteria and are considered as
initial conditions for our core models.

The Bonnor-Ebert (1956, 1955) critical mass for unmagnetized cores bounded by surface pressure $P_s$ is given by
\be
M_{B.E.}=1.18\frac{\sigma^4}{G^{3/2} P_s^{1/2}}.
\label{eq:Mbe}
\ee
Although the ordered magnetic field in our models modifies the critical mass from its
hydrostatic value, we nevertheless find $M_{B.E.}$ to be a useful fiducial mass scale to compare with our magnetized models.
We find that an equilibrium state is likely to be found if the mass of a core
is chosen to be comparable to, but generally not in excess of the Bonnor-Ebert critical mass for non-magnetic cores.
Therefore, we attempt to compute core models with masses ranging from 0.3 to 0.8 (in steps of 0.1)
times $M_{B.E.}$ for each of the 71 filaments chosen as possible initial states.
Models that are more massive than $\sim 0.8~\Mbe$ generally
do not converge because the central density grows without bound until the numerical calculation is halted.
Models less massive than $\sim 0.3~\Mbe$ may converge numerically, but they do not meet our selection criterion that the central density of
the core should be at least that of the initial filament (See Section \ref{sec:numerical}).
Even within our chosen mass range, only 152 of the 426 attempted models ($\sim 1/3$)
converged and satisfied our condition on the central density.

\begin{figure}[h]
\begin{minipage}{0.45\textwidth}
\psfig{file=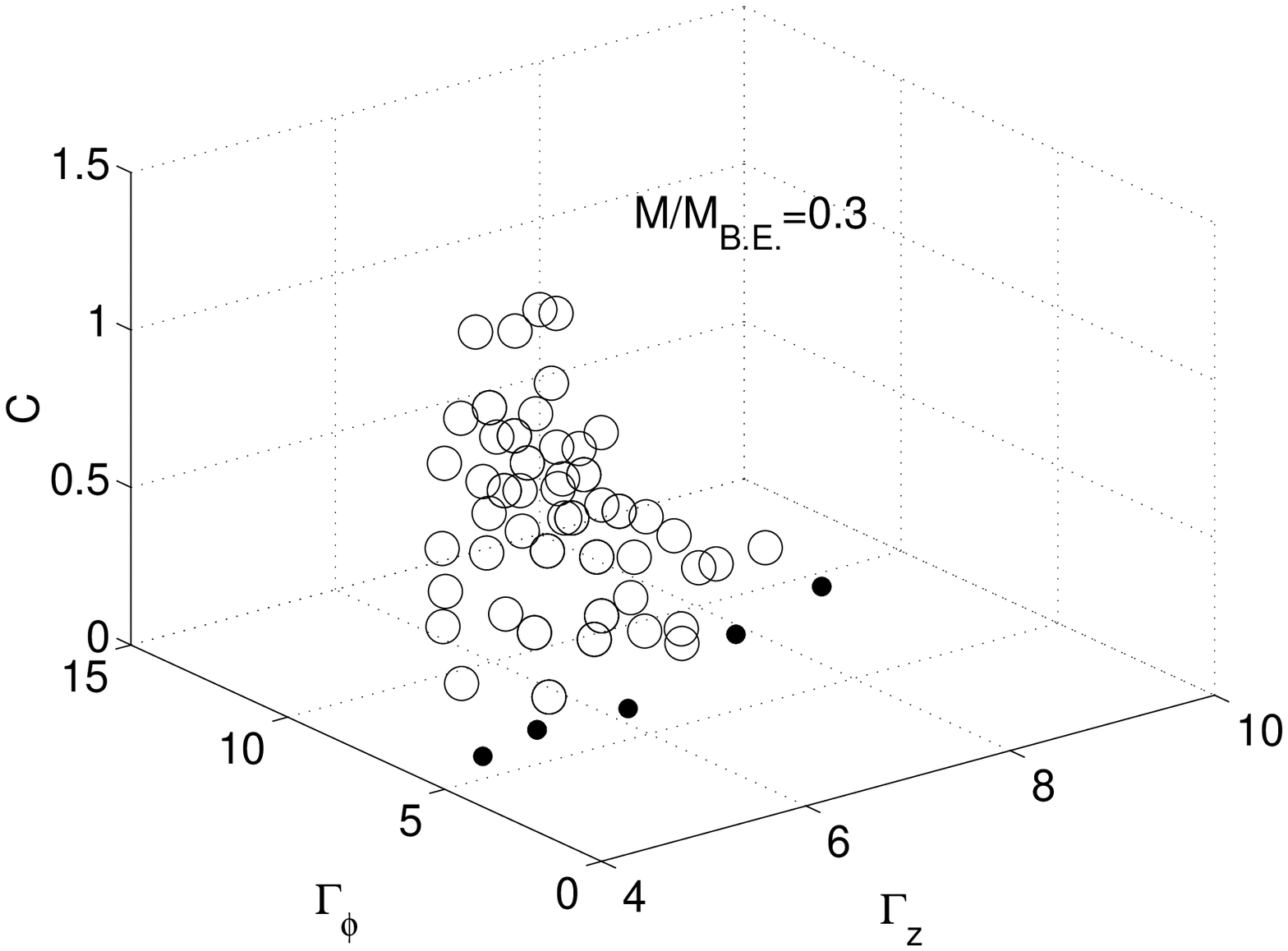,width=\linewidth}
\end{minipage}
\begin{minipage}{0.45\textwidth}
\psfig{file=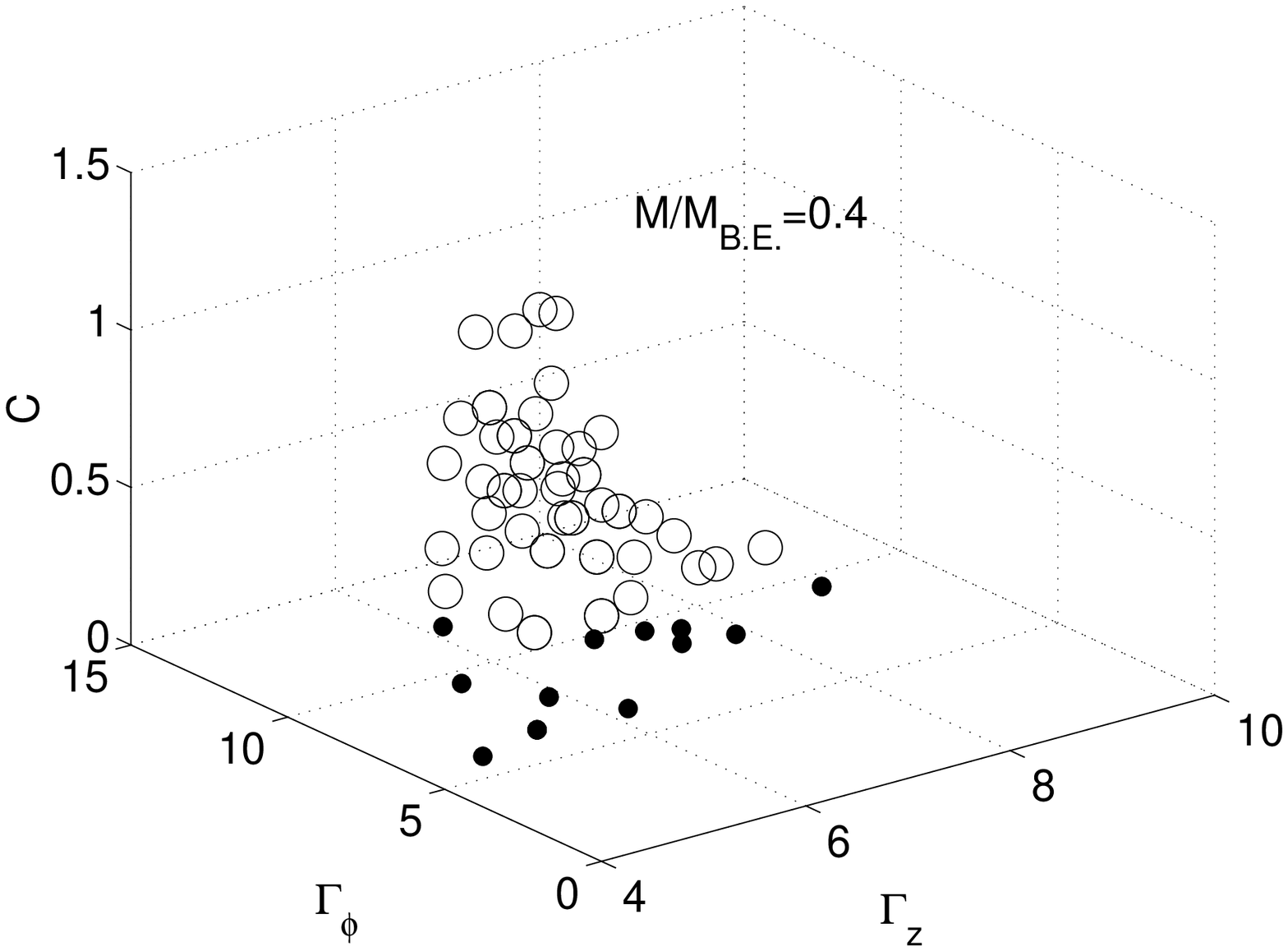,width=\linewidth}
\end{minipage}

\begin{minipage}{0.45\textwidth}
\psfig{file=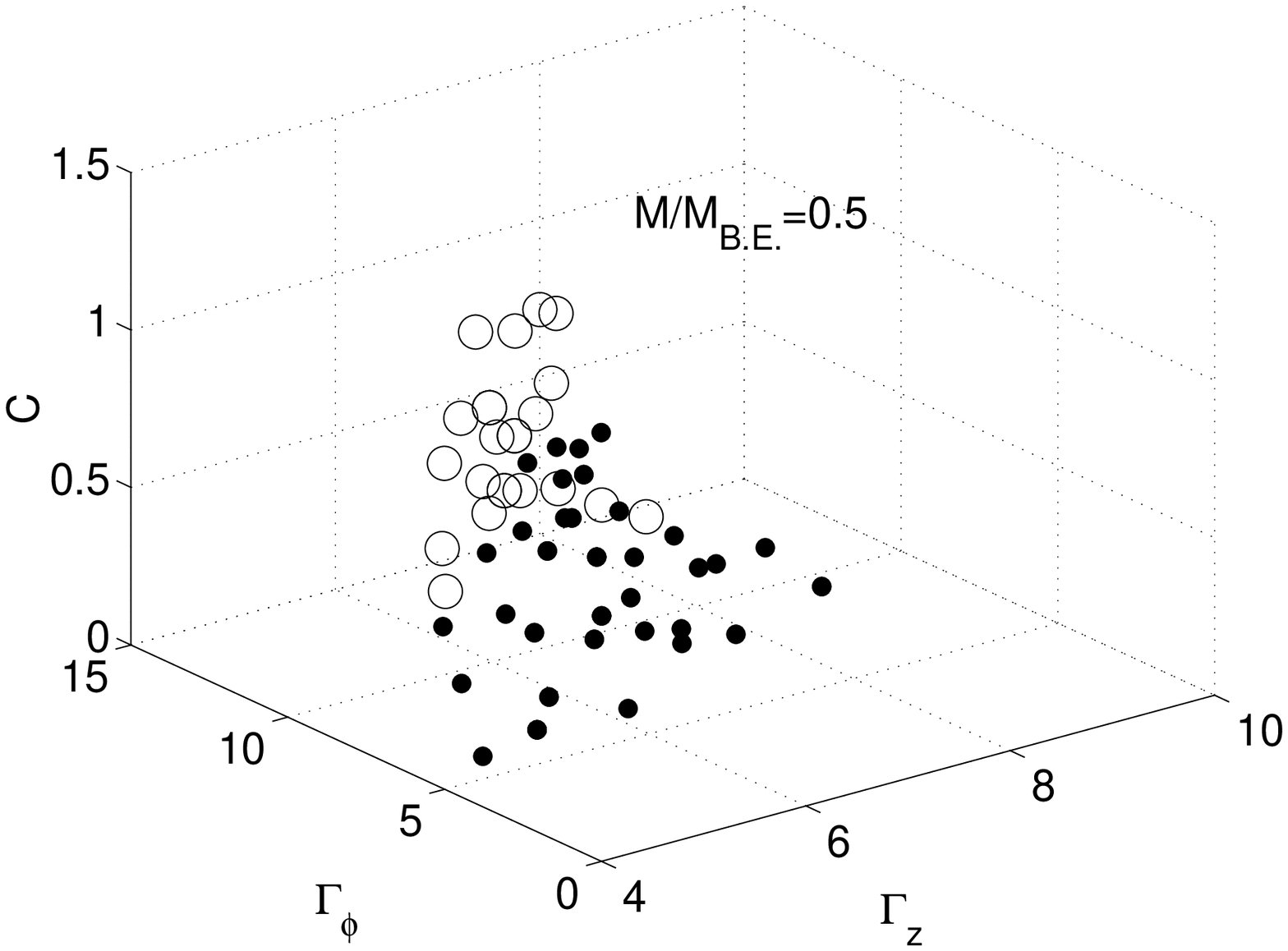,width=\linewidth}
\end{minipage}
\begin{minipage}{0.45\textwidth}
\psfig{file=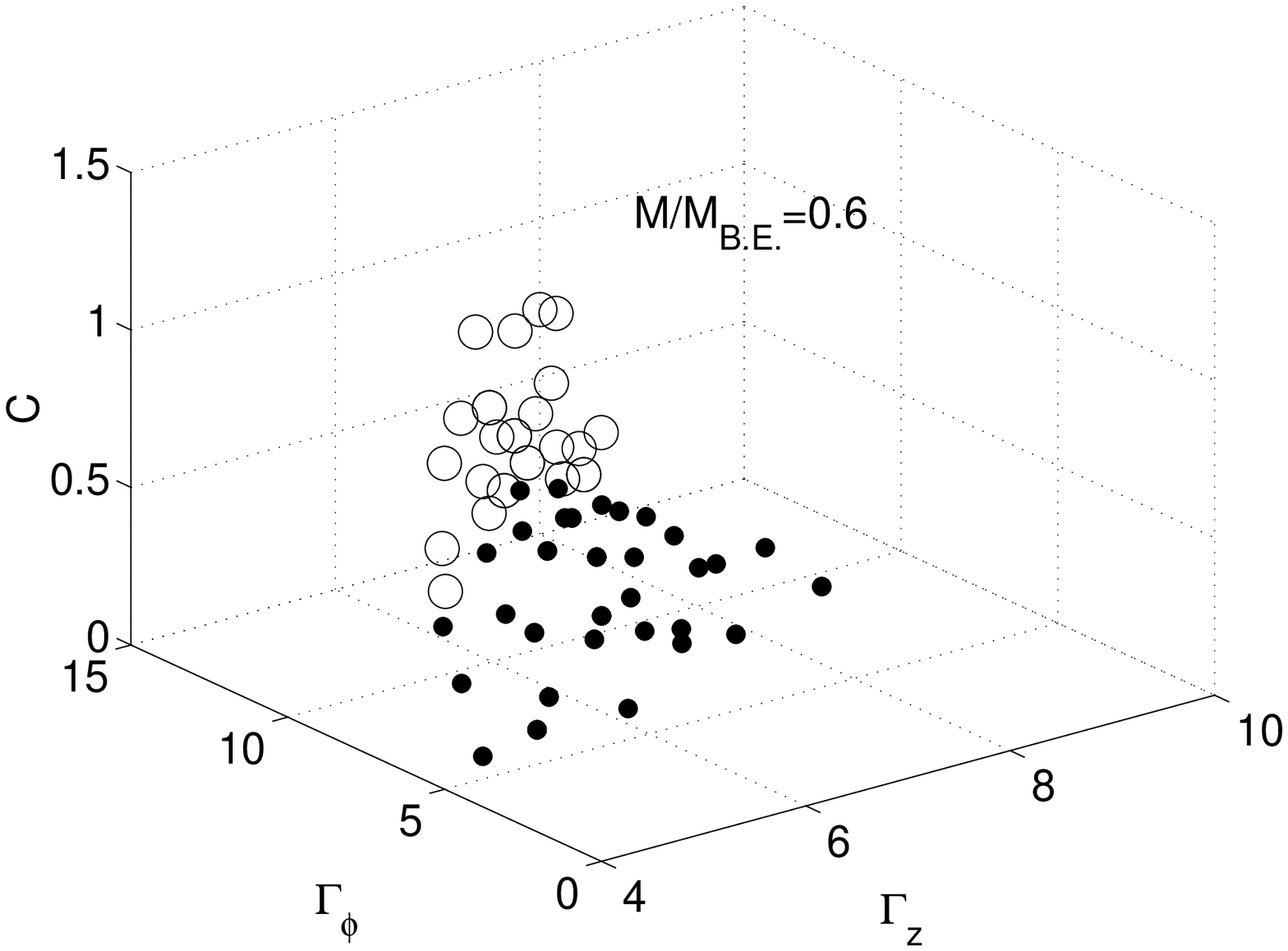,width=\linewidth}
\end{minipage}

\begin{minipage}{0.45\textwidth}
\psfig{file=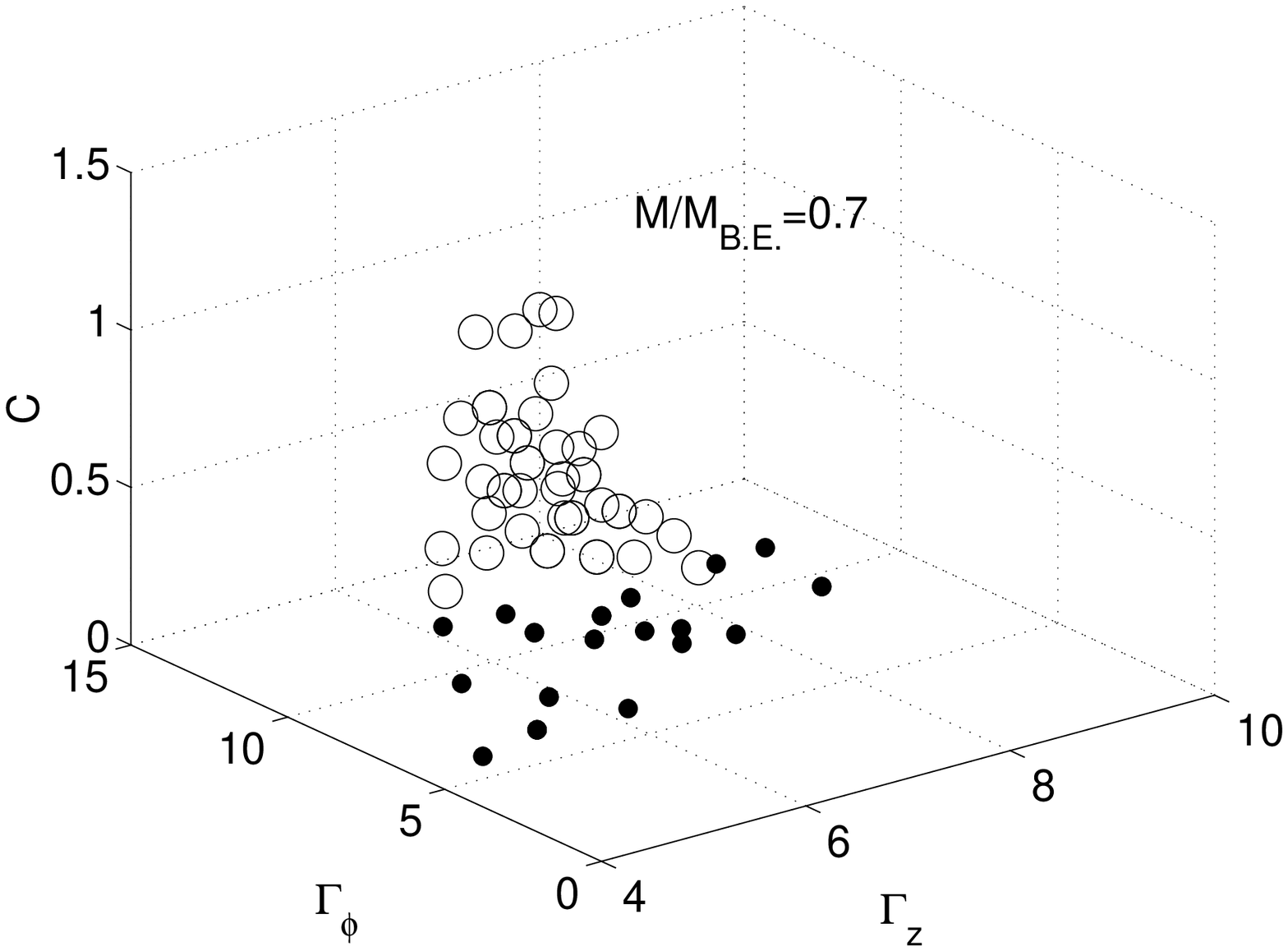,width=\linewidth}
\end{minipage}
\begin{minipage}{0.45\textwidth}
\psfig{file=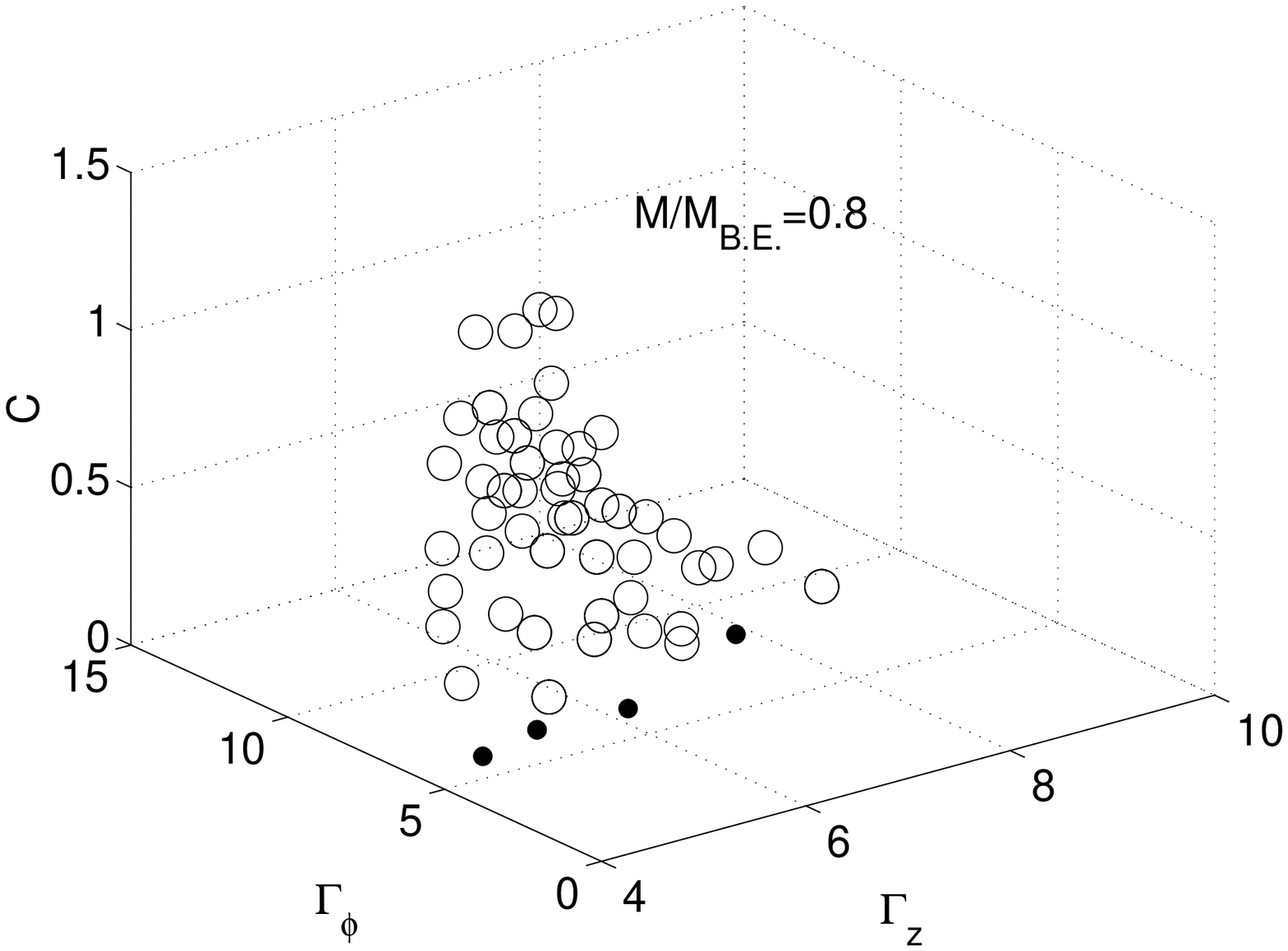,width=\linewidth}
\end{minipage}
\caption{We show the regions of parameter space where our models converge (black dots) or fail to converge (open circles).
Each attempted solution is consistent with a parent filament obeying the observational constraints established in FP1
and having rather low fragmentation growth rates, as given by equation \ref{eq:tauf}.}
\label{fig:slices}
\end{figure}

Figure \ref{fig:slices} shows
6 cuts through our parameter space made by holding $M/\Mbe$ constant.
The solid dots represent models that converged, while the open circles represent failed or rejected models.
It is apparent from the figure that most of the converged models have $0.4\le M/\Mbe \le 0.7$, with only 4 models
found with $M/\Mbe=0.8$ and 5 with $M/\Mbe=0.3$.
Within each cut of constant $M/\Mbe$, we find that our converged models 
are separated from those that did not converge by a fairly sharp boundary, with
little intermixing.  This suggests that models that fail to converge likely represent genuine non-equilibrium or unstable states.
Models do not converge when $M\appgeq 0.8~M_{B.E.}$ because the Bonnor-Ebert critical mass is
reduced in our model by the toroidal field, which works in concert
with gravity and surface pressure in squeezing the cores.  This result is in agreement with Tomisaka (1991) and
Habe et al. (1991).
We also find that $\Gphi$ never exceeds $\Gz$ by more than a factor of $\sim 1.7$ for any
converged model.  This suggests that MHD instabilities might be preventing the convergence of models
with very strong toroidal fields, since we found in FP2 that MHD sausage instabilities were triggered in filaments
when $\Gphi/\Gz\appgeq 2$ (See also discussion in McKee 1999).

\subsection{A Gallery of Models}
\label{sec:gallery}
\begin{figure}[ht]
\begin{minipage}{0.32\textwidth}
\psfig{file=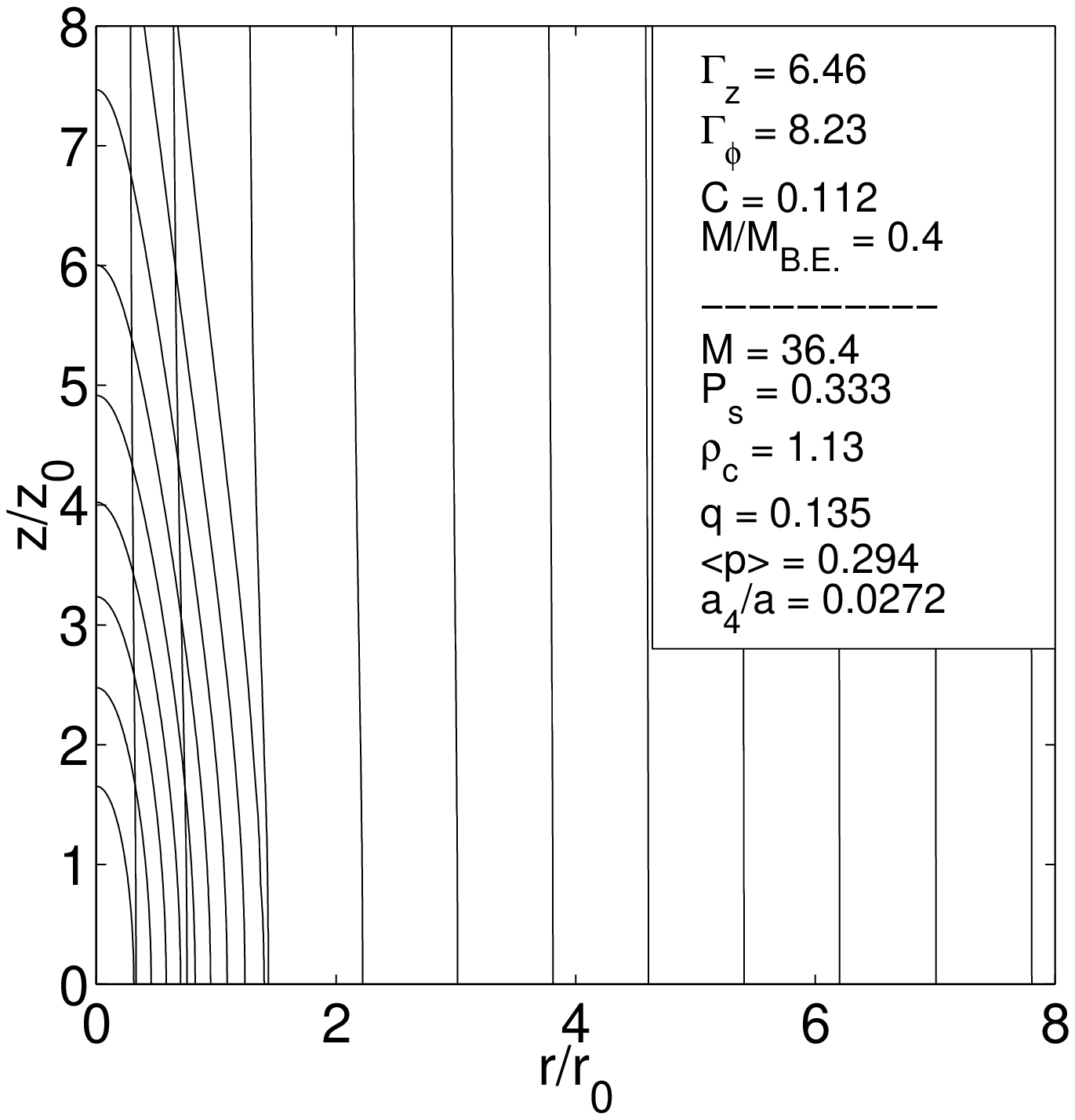,width=\textwidth}
\end{minipage}
\begin{minipage}{0.32\textwidth}
\psfig{file=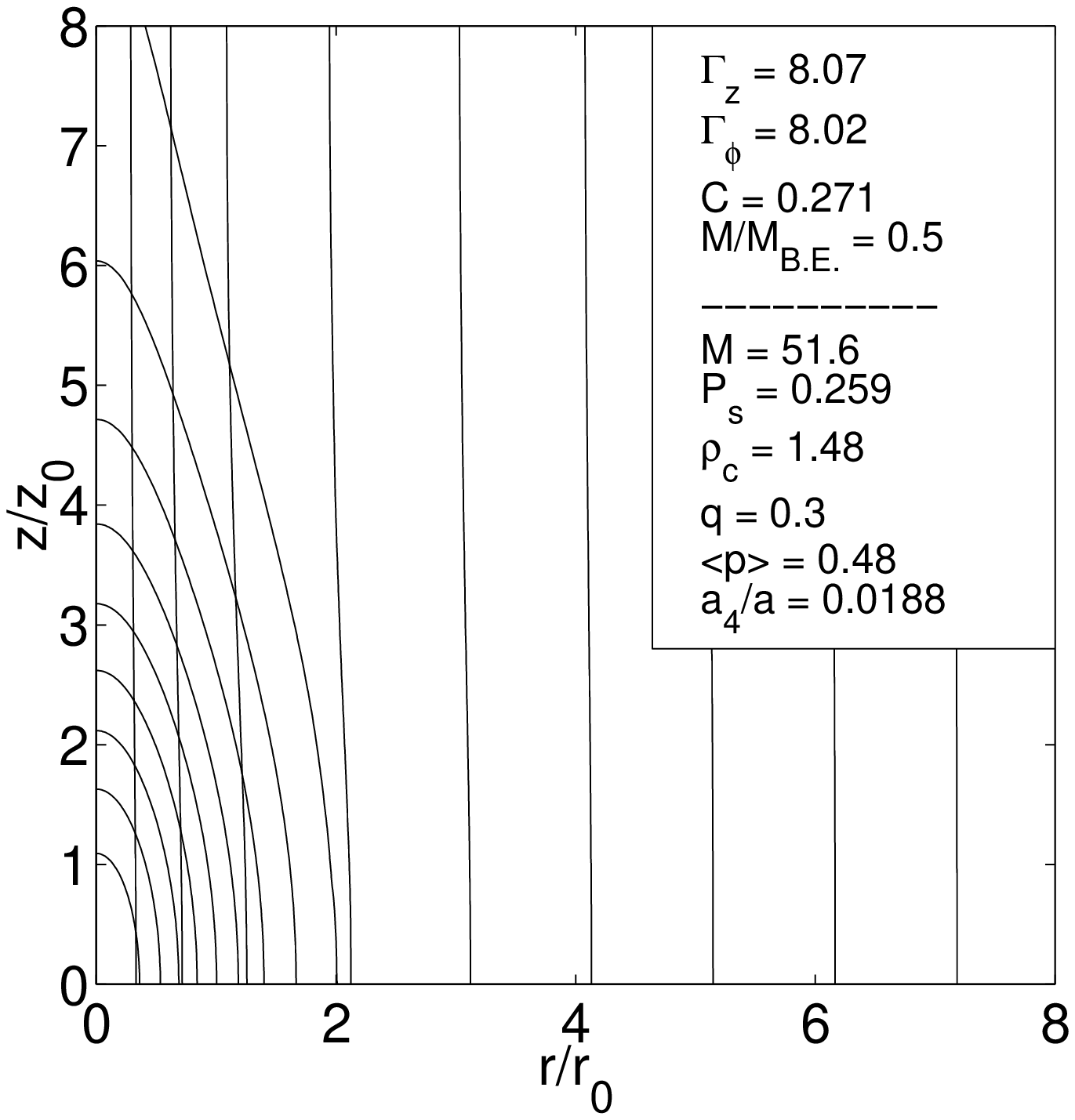,width=\textwidth}
\end{minipage}
\begin{minipage}{0.32\textwidth}
\psfig{file=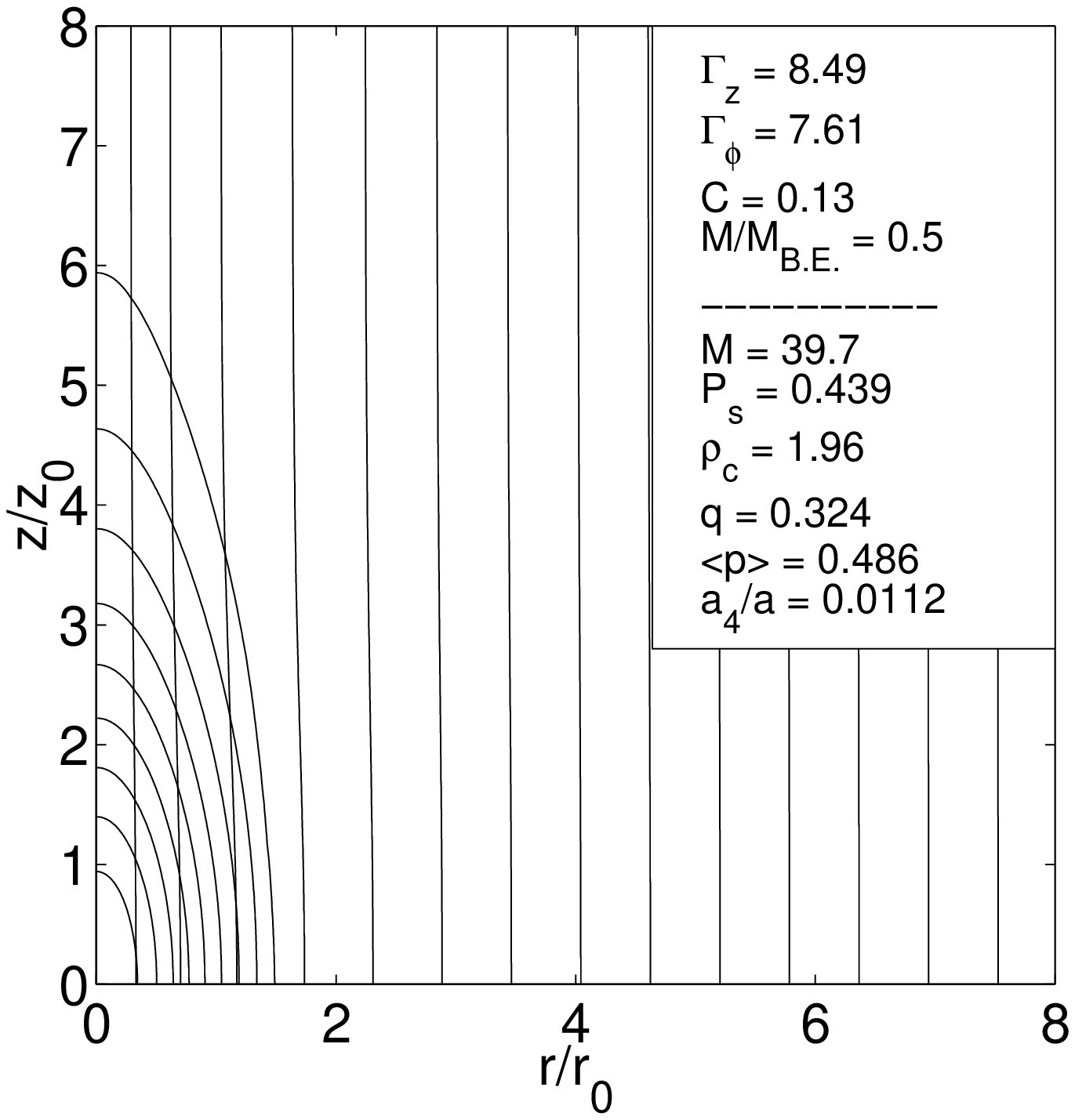,width=\textwidth}
\end{minipage}

\begin{minipage}{0.32\textwidth}
\psfig{file=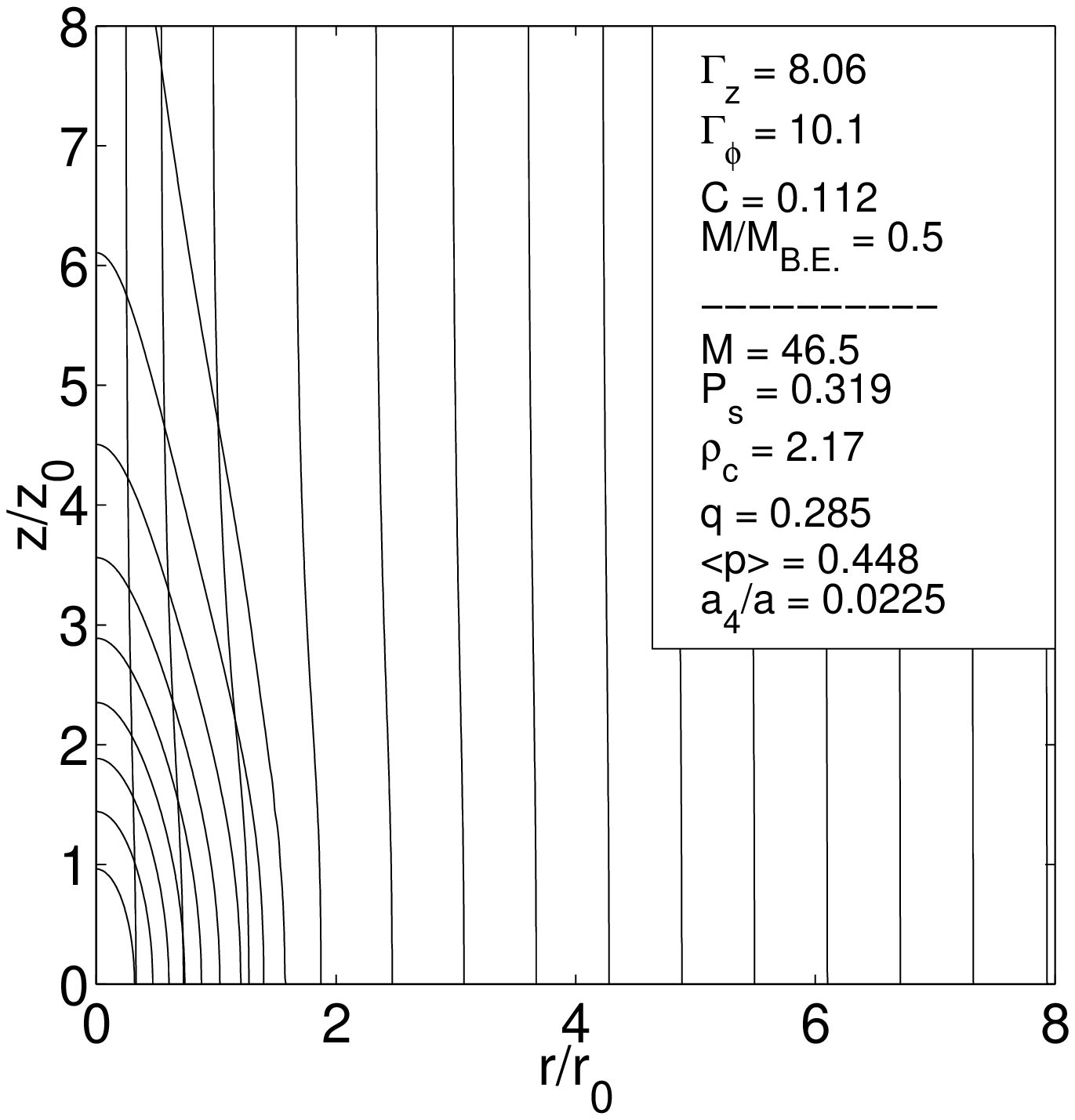,width=\textwidth}
\end{minipage}
\begin{minipage}{0.32\textwidth}
\psfig{file=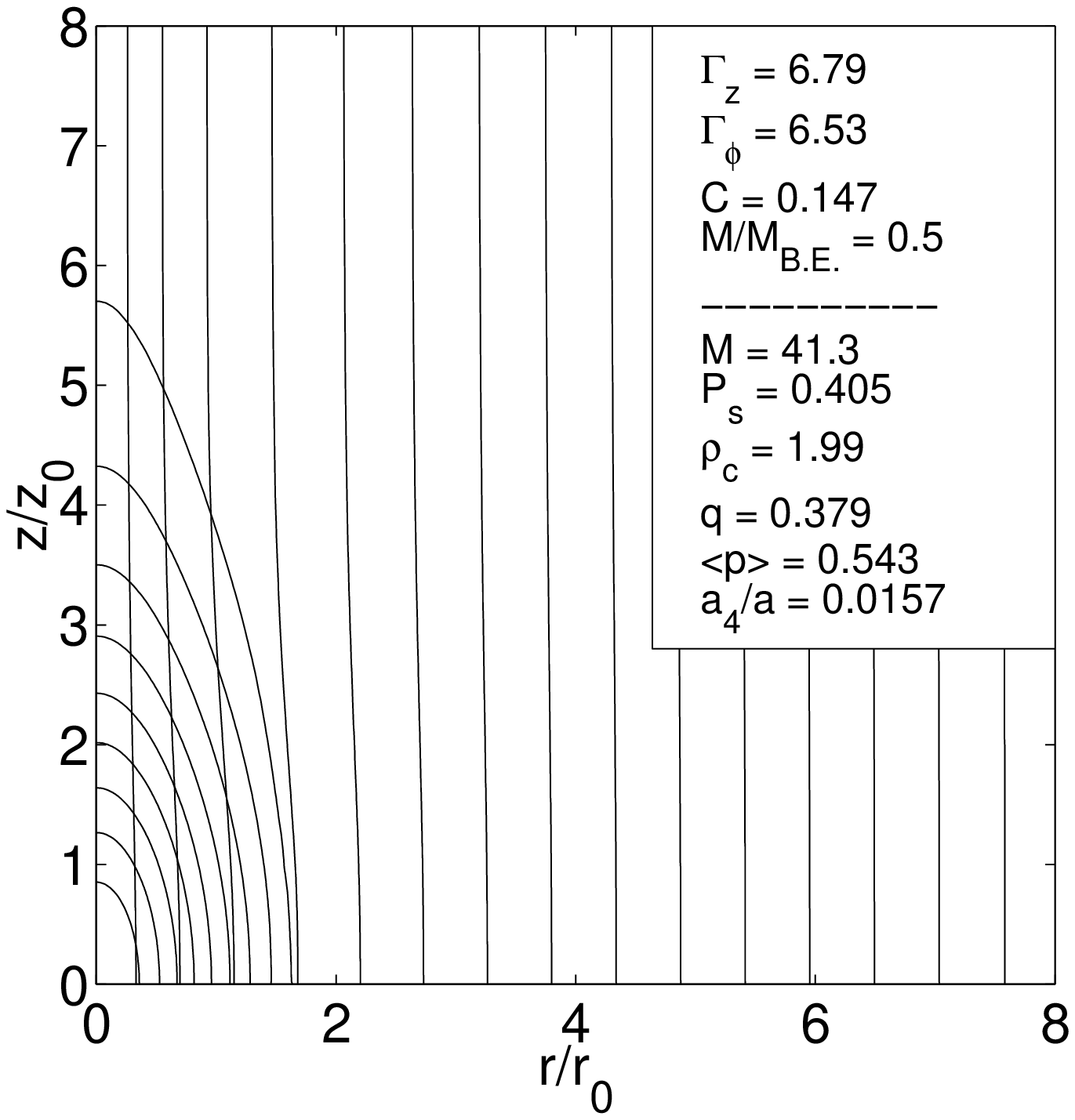,width=\textwidth}
\end{minipage}
\begin{minipage}{0.32\textwidth}
\psfig{file=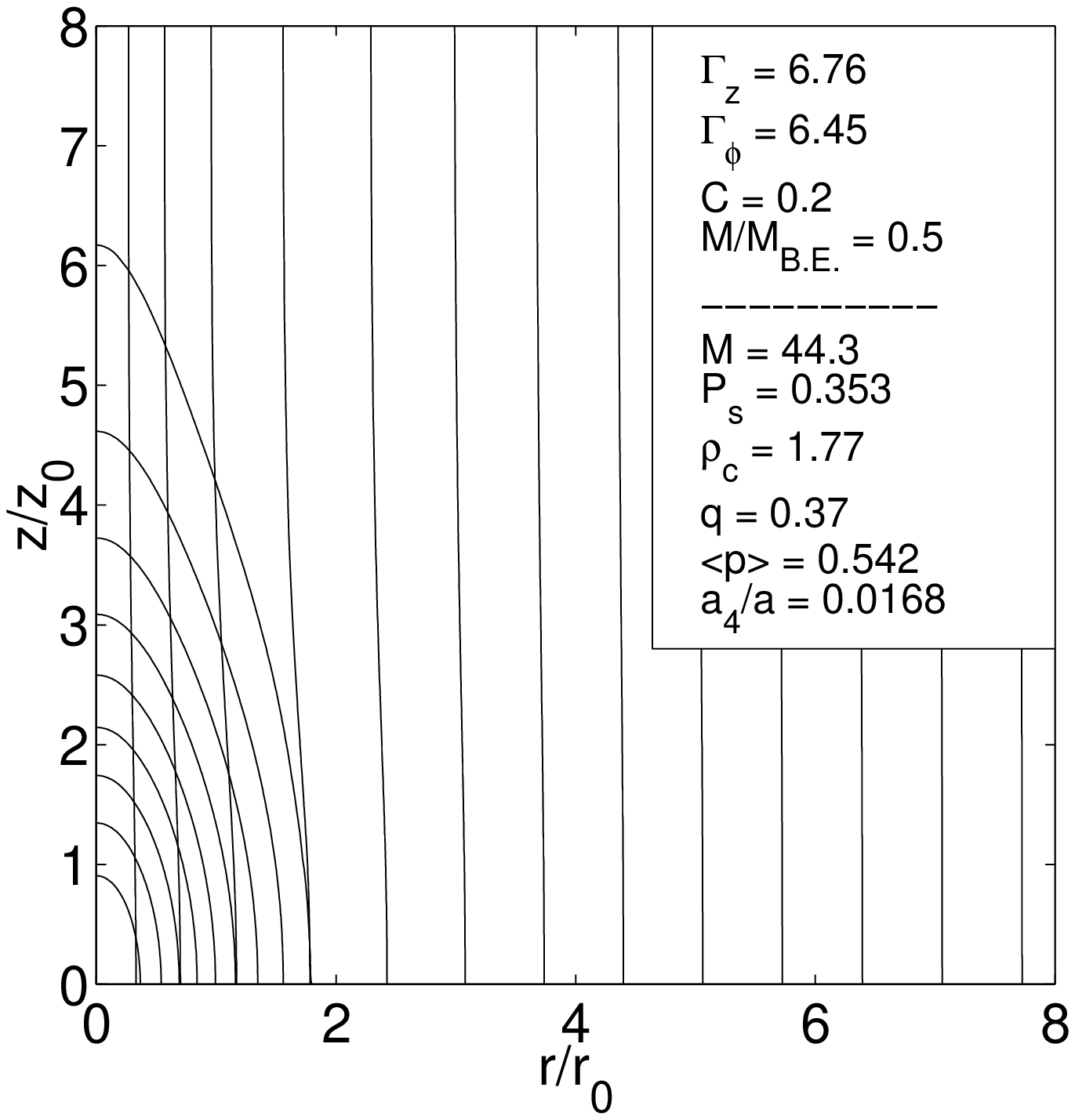,width=\textwidth}
\end{minipage}

\begin{minipage}{0.32\textwidth}
\psfig{file=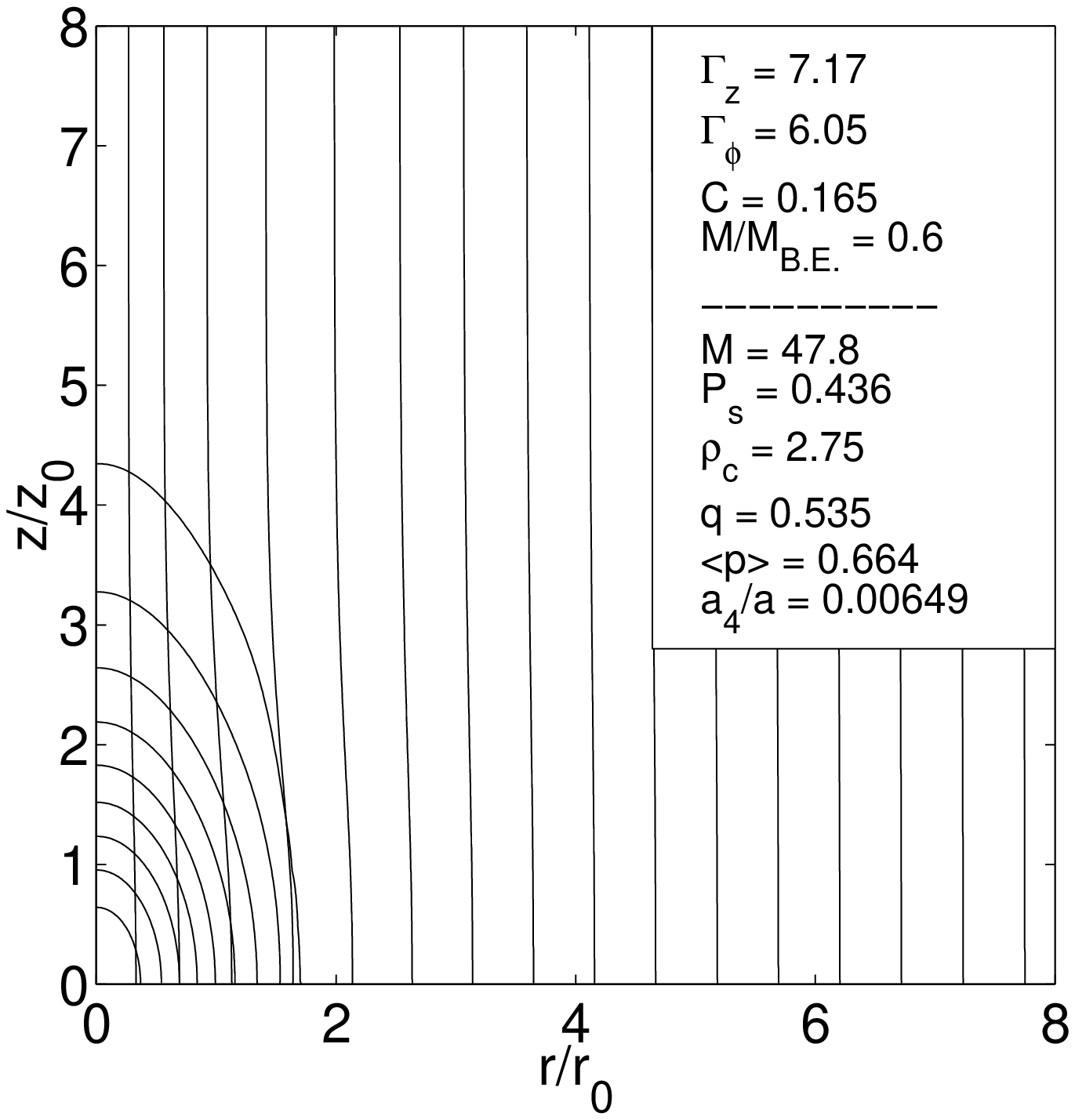,width=\textwidth}
\end{minipage}
\begin{minipage}{0.32\textwidth}
\psfig{file=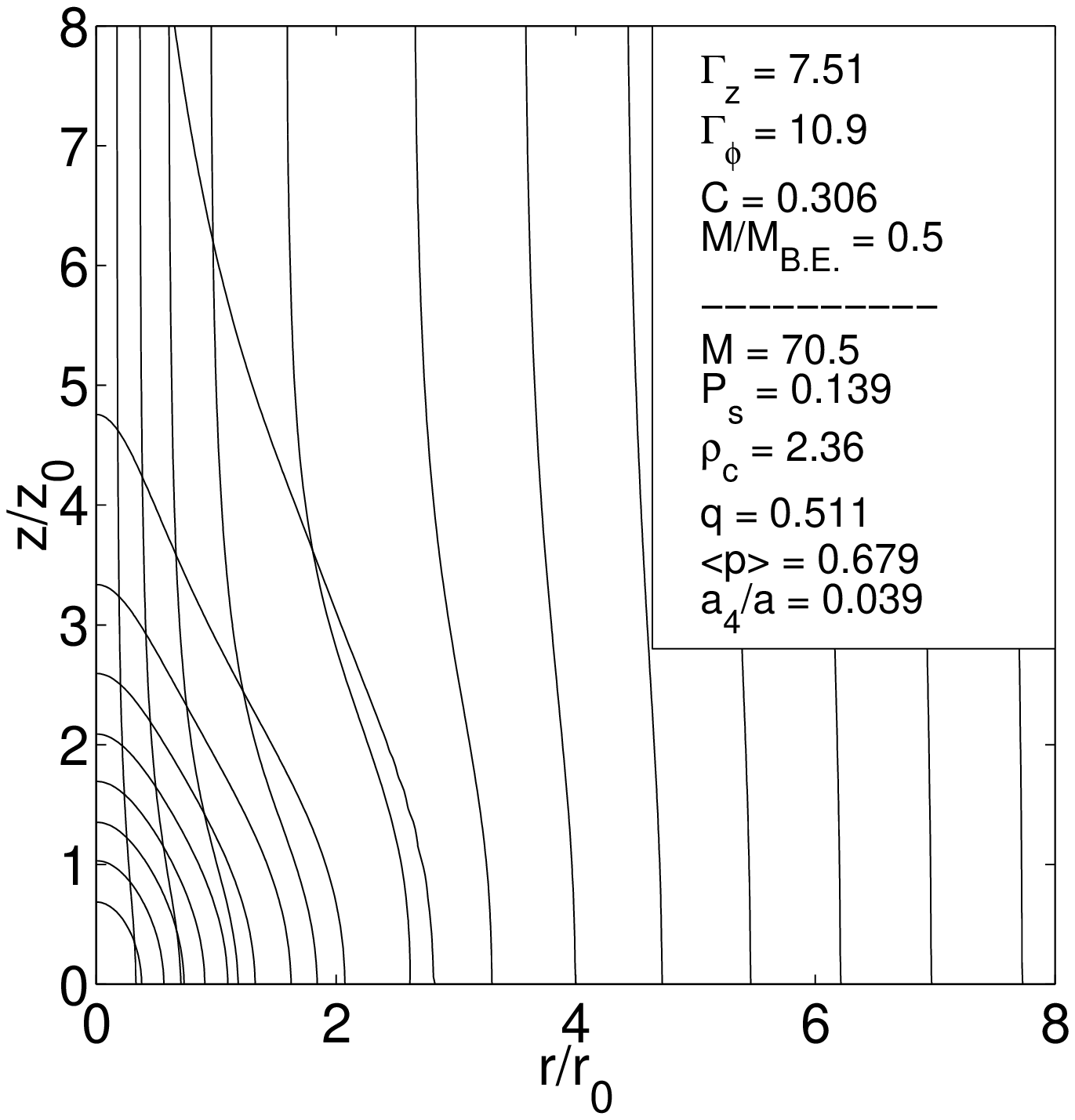,width=\textwidth}
\end{minipage}
\begin{minipage}{0.32\textwidth}
\psfig{file=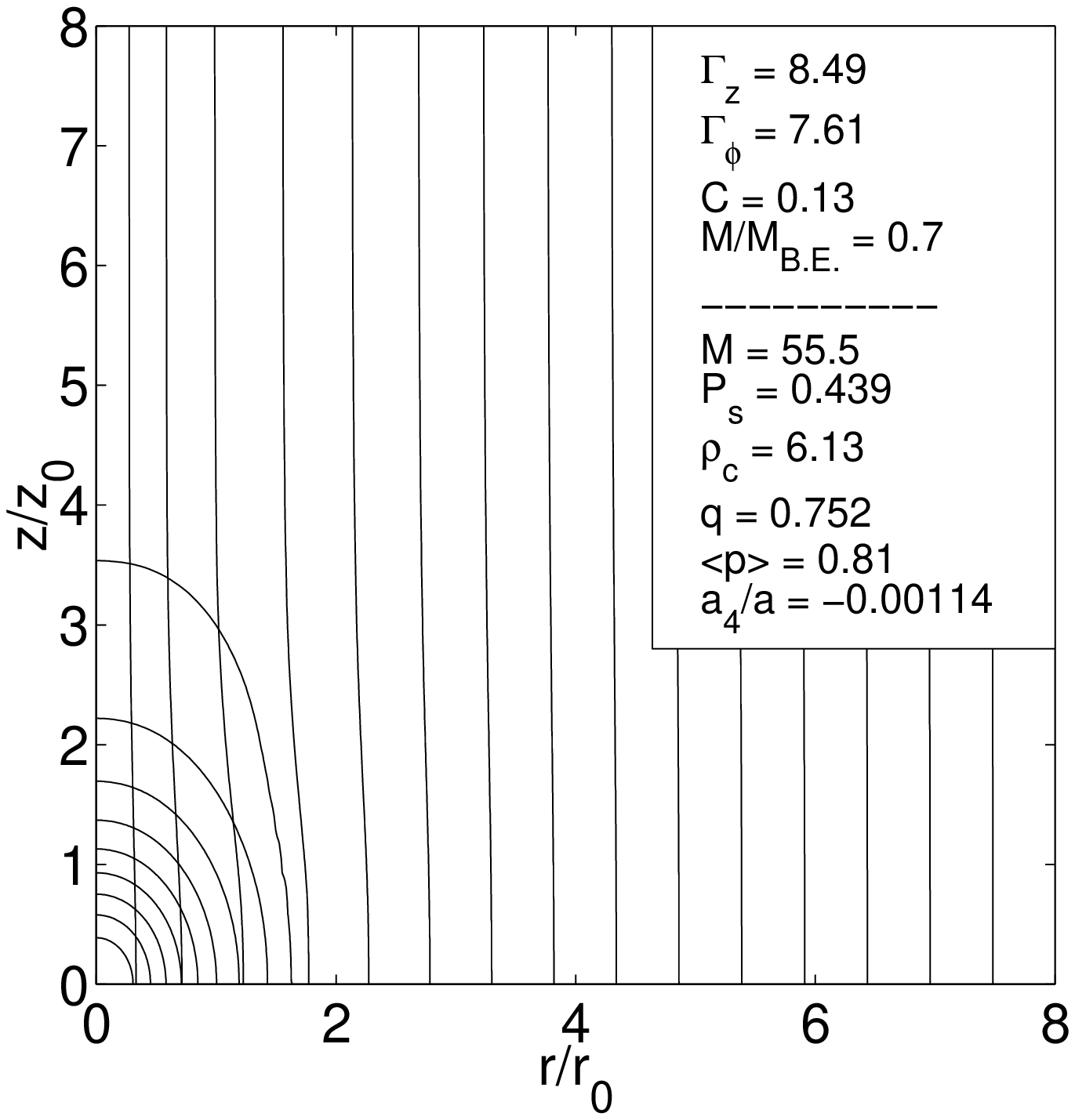,width=\textwidth}
\end{minipage}

\caption{We show density contours and poloidal field lines for a set of representative models to illustrate the range
of equilibria that we obtain.  Models are shown in order of increasing $q$.}
\label{fig:gallery}
\end{figure}

We have computed a total of 152 models scattered throughout our parameter space.
Figure \ref{fig:gallery} shows a gallery of representative models, with a range of intrinsic
axis ratios $q$ ranging from 0.2 to 0.8, shown in order of increasing mass.
The density contours are linearly spaced from the central density (given in each panel) to the 
density at the surface of pressure truncation.  
We find that the density contours near the centre of our models are nearly circular, but become increasingly
elongated towards the surface.
This is explained by the dominance of gravity near the centre, and the increasing 
importance of the toroidal magnetic field in the outer regions.
The density contours in our models are therefore somewhat more complex than in the empirical model
used by Myers et al. (1991) and Ryden (1996), who assume that the density contours of cores can be described
by nested ellipses of constant axis ratio.  

Our models have a central ``core'' region of nearly uniform
density, surrounded by an outer ``envelope,'' where the density gradually decreases with radius.
Figure \ref{fig:gallery} shows that our models are generally quite truncated
by the external pressure, typically extending to only $\sim 2 r_0$ in cylindrical radius.  
Therefore, they are truncated before they attain the $r^{-2}$ density profile (or any other power law) that characterizes
the outer envelope of a Bonnor-Ebert sphere.

For each of our models, we have computed the intrinsic axis ratio $q$ by calculating the best-fitting
ellipse (in a least squares sense) to the half maximum density contour.  We have also calculated the mean projected axis ratio
$\pave$ for each model using the procedure outlined in Appendix A.  (Essentially, $\pave$ is the expectation value of the projected
axis ratio of the half maximum surface density contour for a core oriented at a random inclination angle relative to the plane
of the sky.)

We show histograms of the intrinsic and mean projected axis ratios in Figure \ref{fig:histqp} below,
where we find that $0.1\appleq q\appleq 0.9$ and $0.3 \appleq \pave \appleq 1$.
These histograms should not be interpreted as a prediction of the distribution of axis ratios for a sample of cores,
as shown in Ryden (1996) for
example.  What they do show is the distribution of axis ratios that was actually obtained by our sample of converged models,
which provides an indication of the relative ease or difficulty in producing core models with a given axis ratio.
The mean value of $q$ derived from observations is probably between
0.4 and 0.5 (Myers et al. 1991, Ryden 1996), but the predicted distributions are quite broad, containing
significant numbers of cores
that are much more elongated or more nearly spherical (Ryden 1996).  Therefore, Figure \ref{fig:histqp}
shows that our models cover the likely range of axis ratios.

\begin{figure}[ht]
\psfig{file=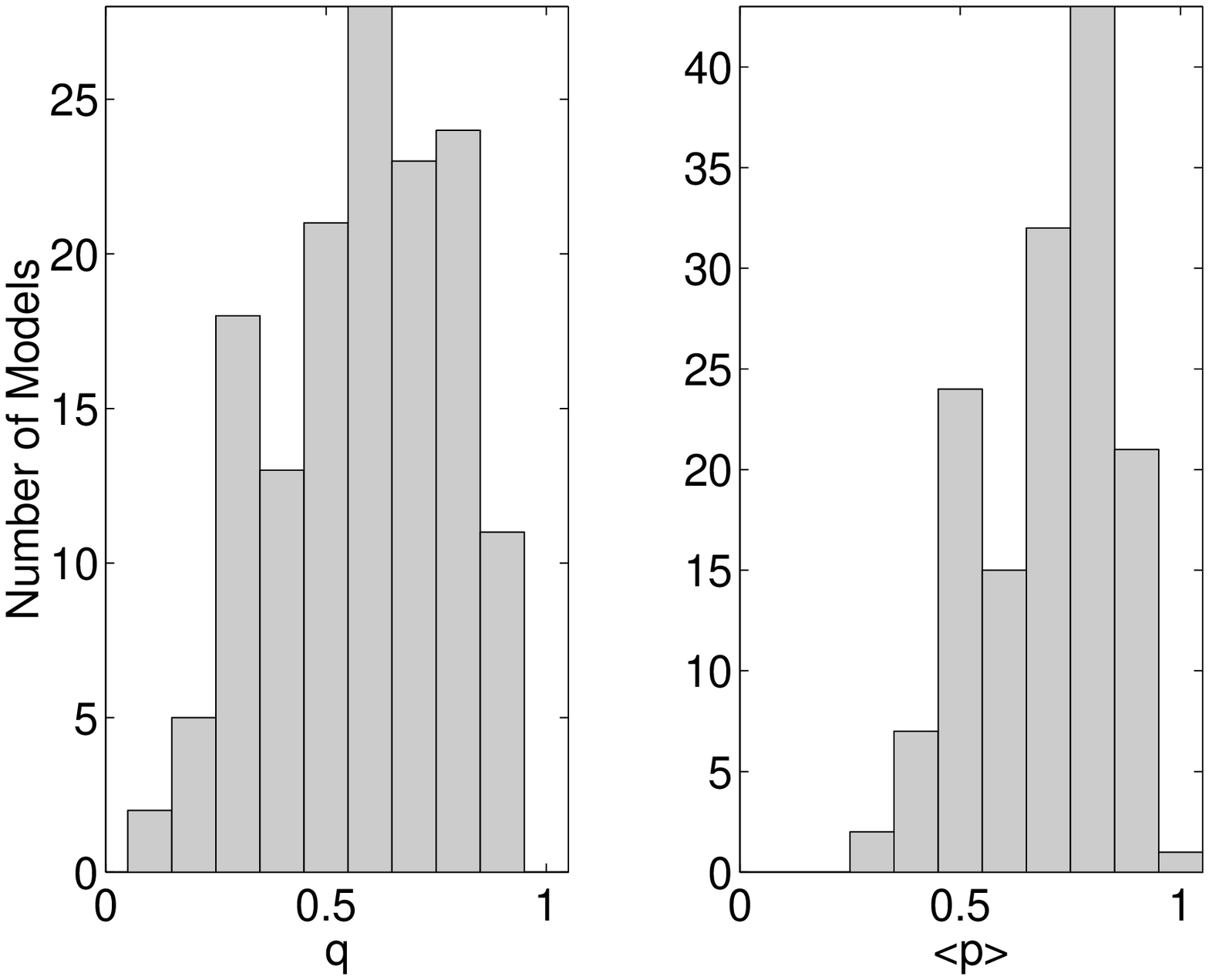,width=\linewidth}
\caption{We show the distribution of intrinsic and mean projected axis ratios obtained for our converged models.
All 152 models are elongated, many with axis ratios in good agreement with the observed values.}
\label{fig:histqp}
\end{figure}

\begin{figure}[ht]
\psfig{file=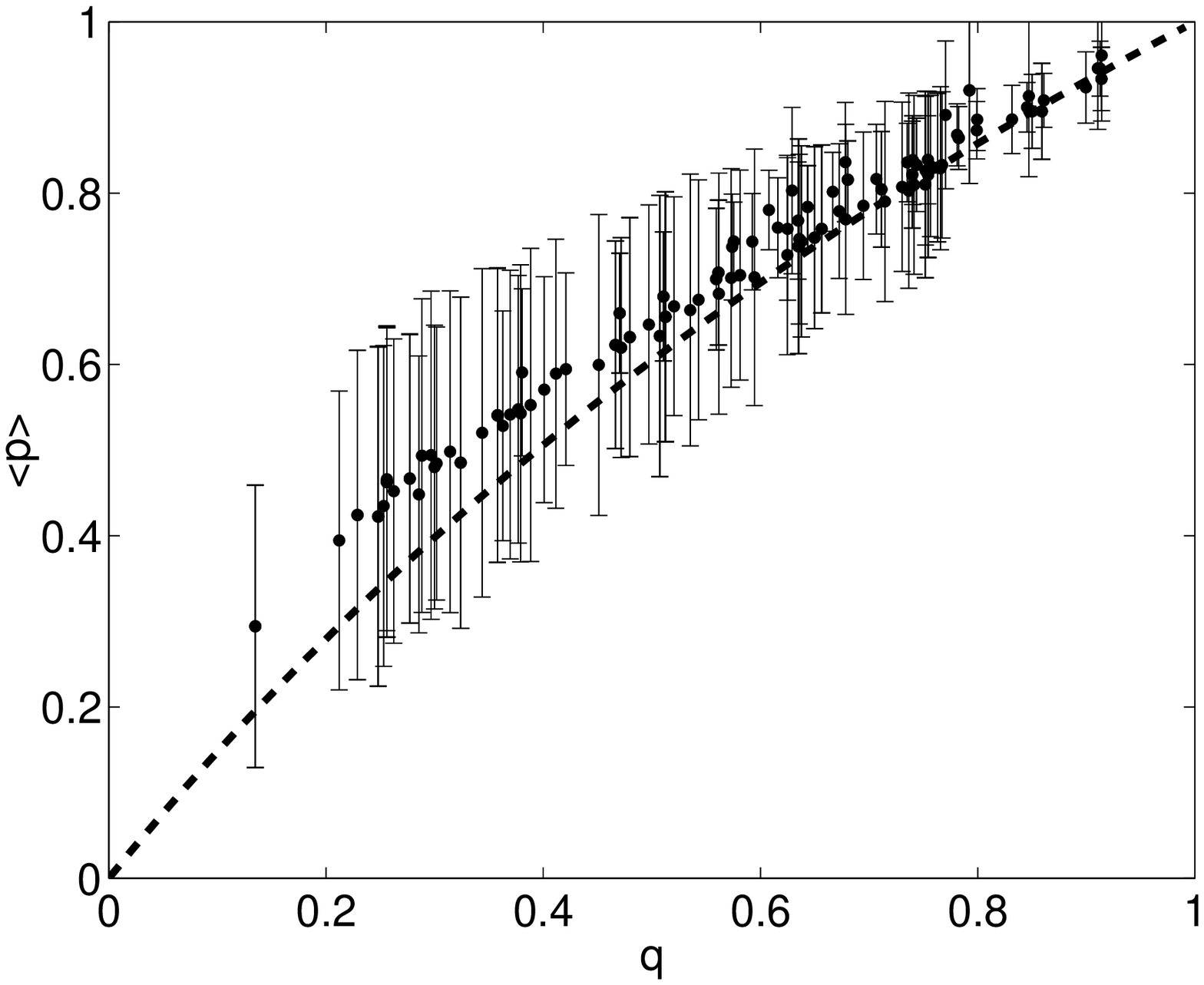,width=\linewidth}
\caption{We show the relationship between the intrinsic and mean projected axis ratios for our model.  We find reasonable
agreement with the formula (dashed line) given by Myers et al. (1991) for an empirical model of prolate cores.  The error bars
represent the standard deviation of $p$ about $\pave$, assuming that cores are oriented randomly on the sky.}
\label{fig:qp}
\end{figure}

Figure \ref{fig:qp} shows that the relationship between $\pave$ and $q$, numerically determined by our model,
is in reasonable agreement with the analytical relation (dashed line) used by Myers et al. (1991, their equation 1),
even though the internal structure of our models is quite different from theirs.
The relation between $q$ and $\pave$ is clearly not very sensitive to the detailed internal structure of cores.
The error bars represent the standard deviation in $\pave$ for
a sample of cores with axis ratio $q$, oriented randomly relative to the observer.

We also estimate the degree to which the half maximum density and surface density contours of our models
deviate from a pure ellipse.
Our method for this analysis is the Fourier method developed by Bender and M\"{o}llenhoff (1987),
who used the method to quantify the ``diskiness'' or
``boxiness'' of elliptical galaxy isophotes.  Their method works by finding the best-fitting (in a least squares sense)
ellipse to an isophote and Fourier analysing the residual.  Their analysis shows that the fourth Fourier
component $a_4$ of the residual dominates the Fourier spectrum, and when normalized with respect to the semimajor axis $a$
provides an excellent indication of the deviation from an ellipse.  For our purposes, $a_4/a>0$ describes density 
or surface density contours that
are more peaked than an ellipse along the axis of symmetry (ie: shaped like a North-American football or lemon), 
while $a_4/a<0$ describes contours that are more flattened than an ellipse.  

\begin{figure}[ht]
\psfig{file=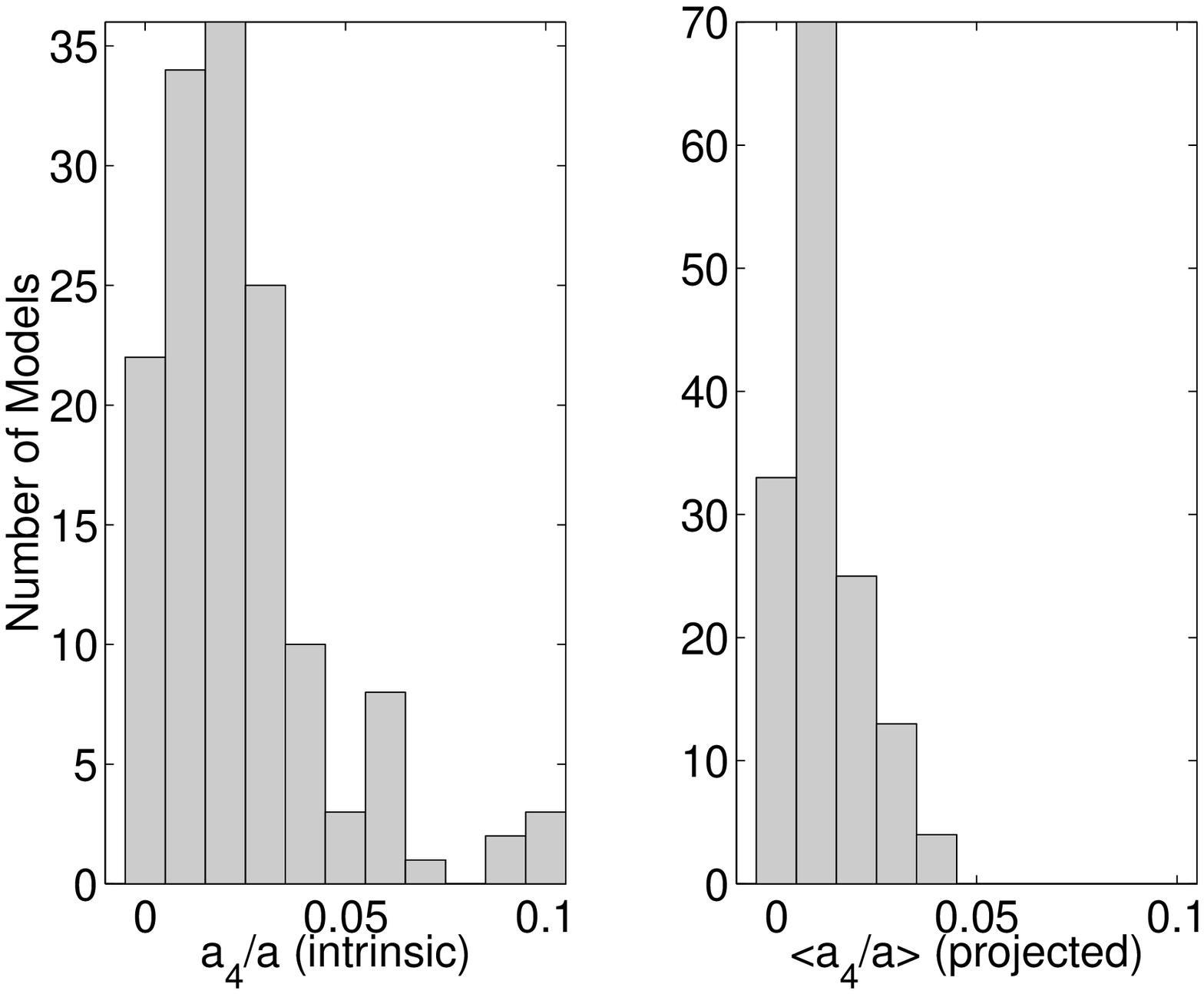,width=\linewidth}
\caption{We show the distribution of the intrinsic and mean projected $a_4/a$ shape 
parameters for our converged models.}
\label{fig:hista4}
\end{figure}

For each of our models, we calculate both an intrinsic $a_4/a$ parameter  
for the half maximum density contour, and a mean projected $\aave$ parameter for the half maximum surface density
contour, averaging over all possible orientations using the procedure described in Appendix \ref{app:proj}.
We show the distributions of these shape parameters for our models in Figure \ref{fig:hista4}.
We find that all of our models have intrinsic and projected shape parameters that are greater than or equal to zero, 
but usually less than $\sim 0.05$.  Therefore, many of our models are slightly football-shaped.
This might provide one way to distinguish observationally between our models and flattened models
threaded by purely poloidal fields (eg. TIN).

We find no clear correlation between $a_4/a$ and $q$.  However, very few elongated models (with
$\pave\appleq 0.6$) are significantly peaked ($\aave \appgeq 0.02)$.
Conversely, those with a high value of $a_4/a$ are never very elongated.

An interesting feature found in many of our models is that the field lines bulge slightly outward around the core
(See Figure \ref{fig:gallery} for several examples.).
This is opposite the hourglass shaped field lines found by Mouschovias (1976b), TIN, TIN88b, TIN89, TIN90, and Tomisaka (1991).
Physically, this field structure is mainly due to the radial pinch of the toroidal field,
which most significantly affects the density and field line structure near the tips of cores (where the symmetry axis intersects
the surface of the core).
The gas pressure is lowest and the gravitational forces are weakest near the tips, so the toroidal field can squeeze in further than in the midplane.

\begin{figure}[ht]
\begin{minipage}{0.45\textwidth}
\psfig{file=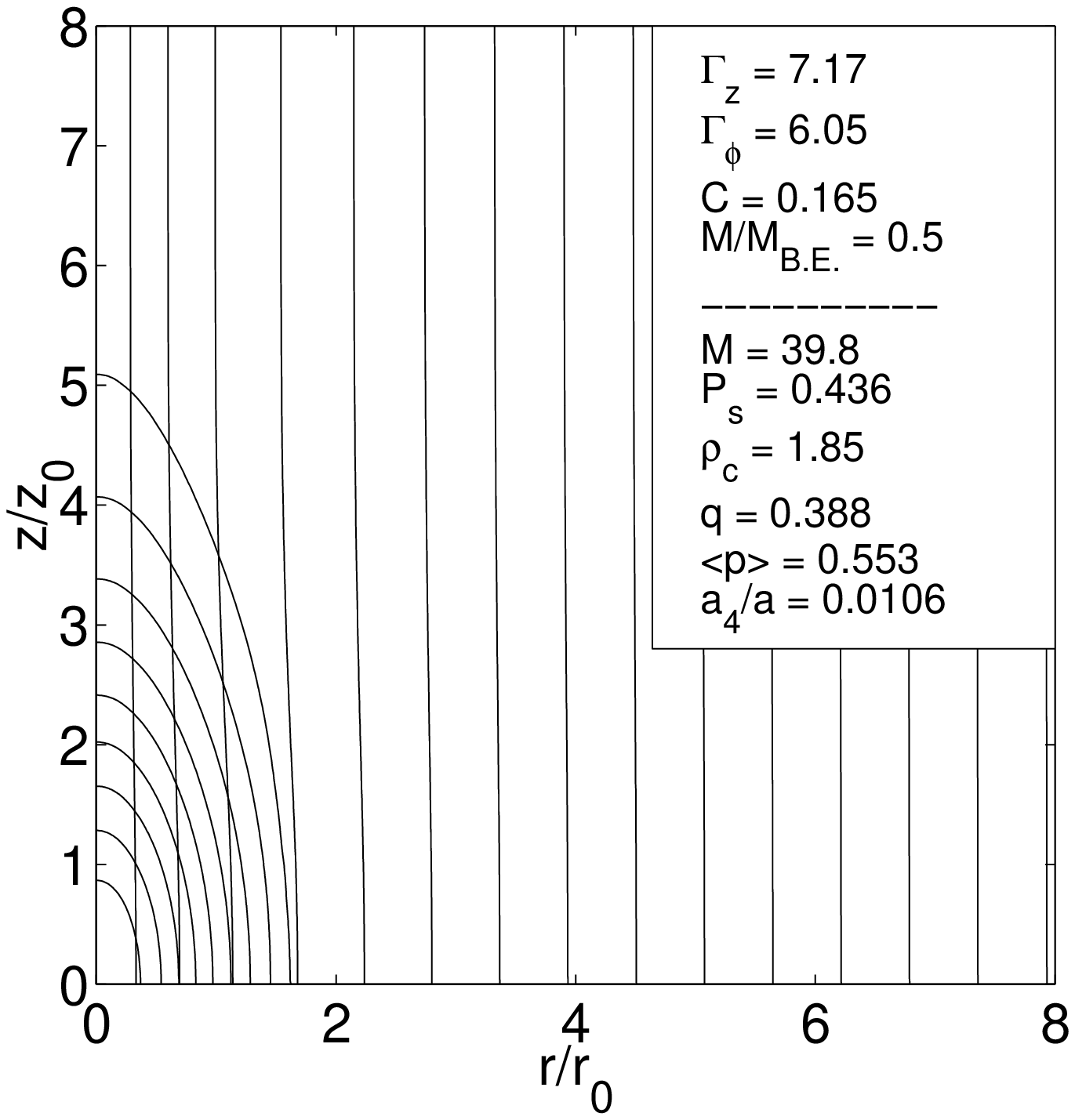,width=\textwidth}
\end{minipage}
\begin{minipage}{0.45\textwidth}
\psfig{file=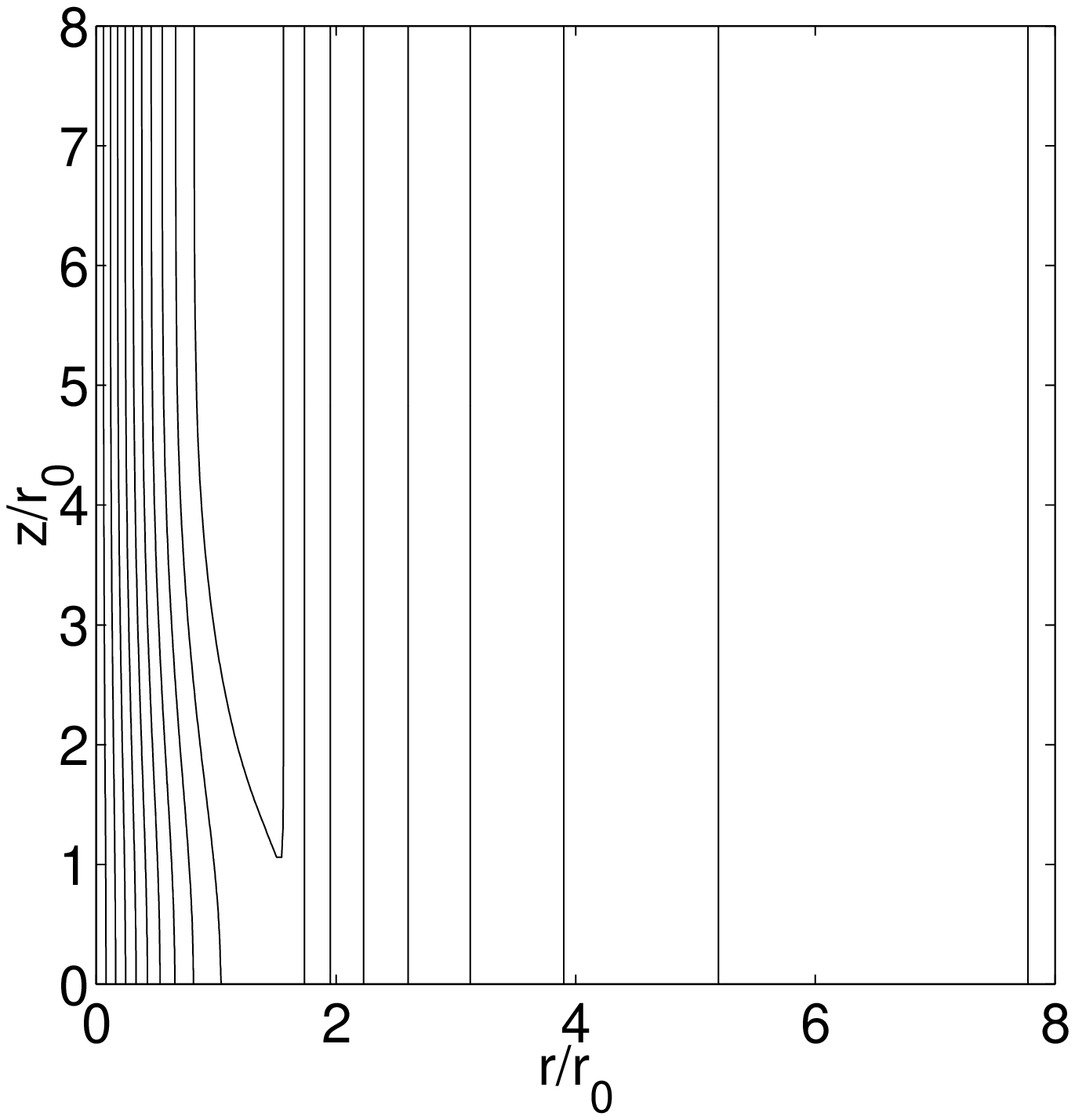,width=\textwidth}
\end{minipage}

\begin{minipage}{0.45\textwidth}
\psfig{file=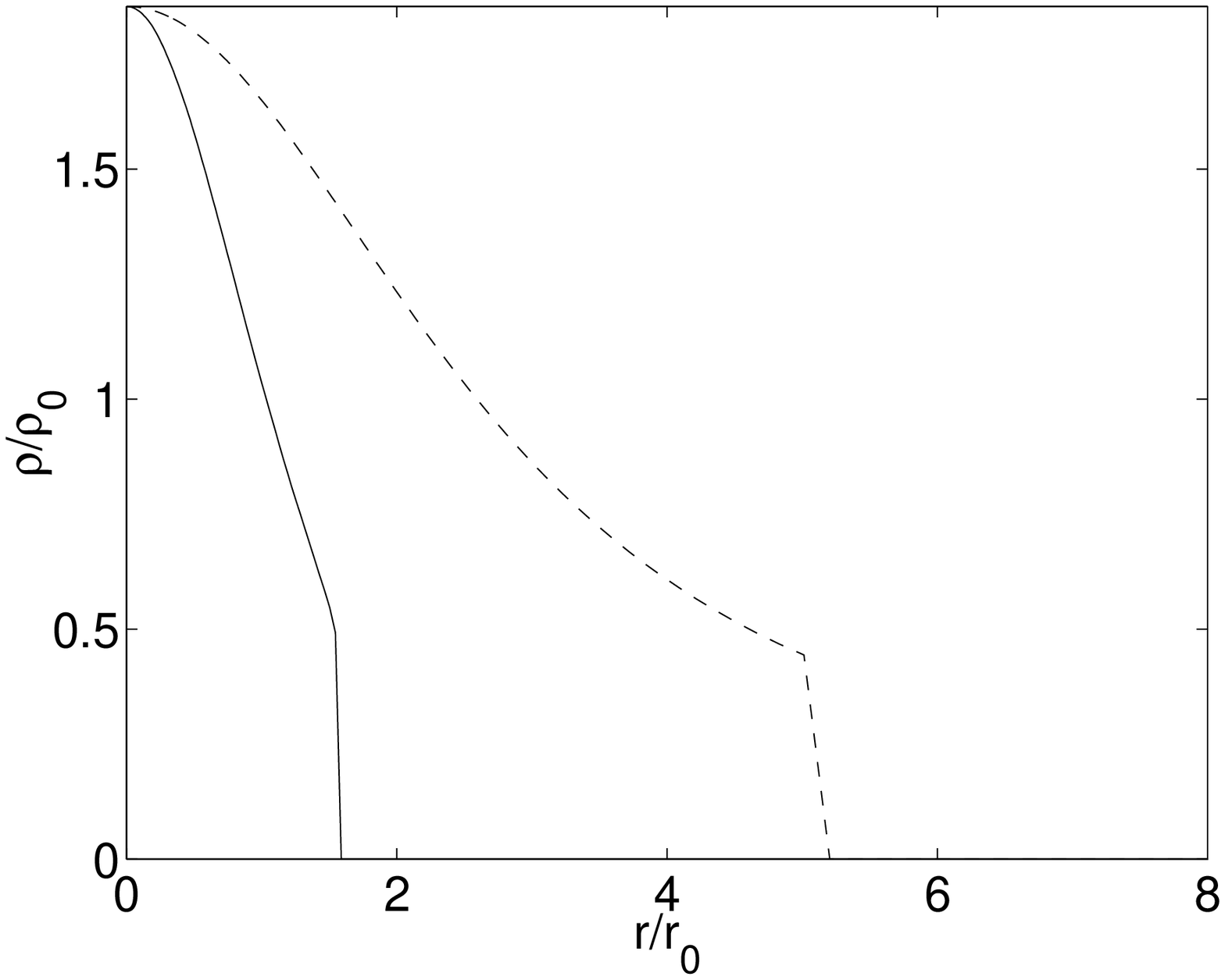,width=\textwidth}
\end{minipage}
\begin{minipage}{0.45\textwidth}
\psfig{file=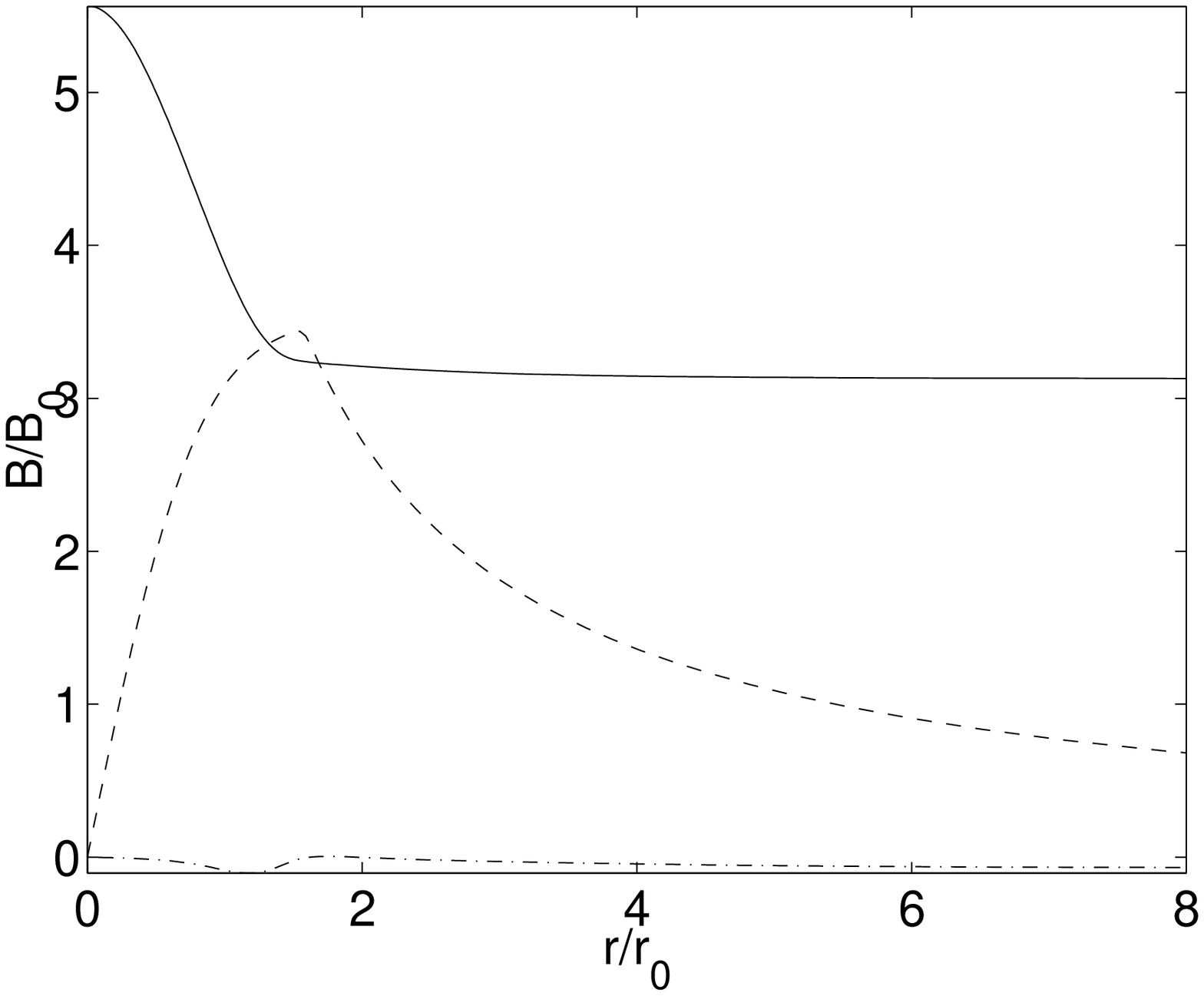,width=\textwidth}
\end{minipage}
\caption{We show the structure of a representative model whose toroidal field is relatively weak compared to its poloidal
field nearly everywhere.  {\bf panel (a)} (upper left): density contours and poloidal field lines.
{\bf panel (b)} (upper right):} Contours of toroidal field intensity.  {\bf panel (c)} (lower left): Density cuts taken radially in
the  midplane (solid line) and along the asymmetry axis (dashed line).  {\bf panel (d)} (lower right): The variation of $\Bz$ (solid
line), $\Bphi$ (dashed line), and $\Br$ (dot-dashed line) taken along a radial cut in the midplane.
\label{fig:sample1}
\end{figure}

We find a continuum of models that range considerably in shape and ellipticity.
We discuss two representative models chosen because of their contrasting behaviour.
The model shown in Figure \ref{fig:sample1} is dominated by the $\Bz$ component of the magnetic
field nearly everywhere.  
The toroidal field vanishes at the axis of symmetry and increases radially outward, becoming comparable to the poloidal field
only very near the edge of the core.  It is strongest near the tips of the cores
rather than in the midplane.  The radial field component is much
smaller than $\Bz$ everywhere within the core.
Many models like the one shown in Figure \ref{fig:sample1}, with a strong
``backbone'' of poloidal flux, are very elongated.  We find models similar to the one shown, but with
intrinsic axis ratios as low as $\sim 0.1$ and mean projected axis ratios as low as $\sim 0.3$.
The density contours are usually quite nearly elliptical when a strong poloidal field is present,
with $0< a_4/a\appleq 0.03$ in most cases.

\begin{figure}[ht]
\begin{minipage}{0.45\textwidth}
\psfig{file=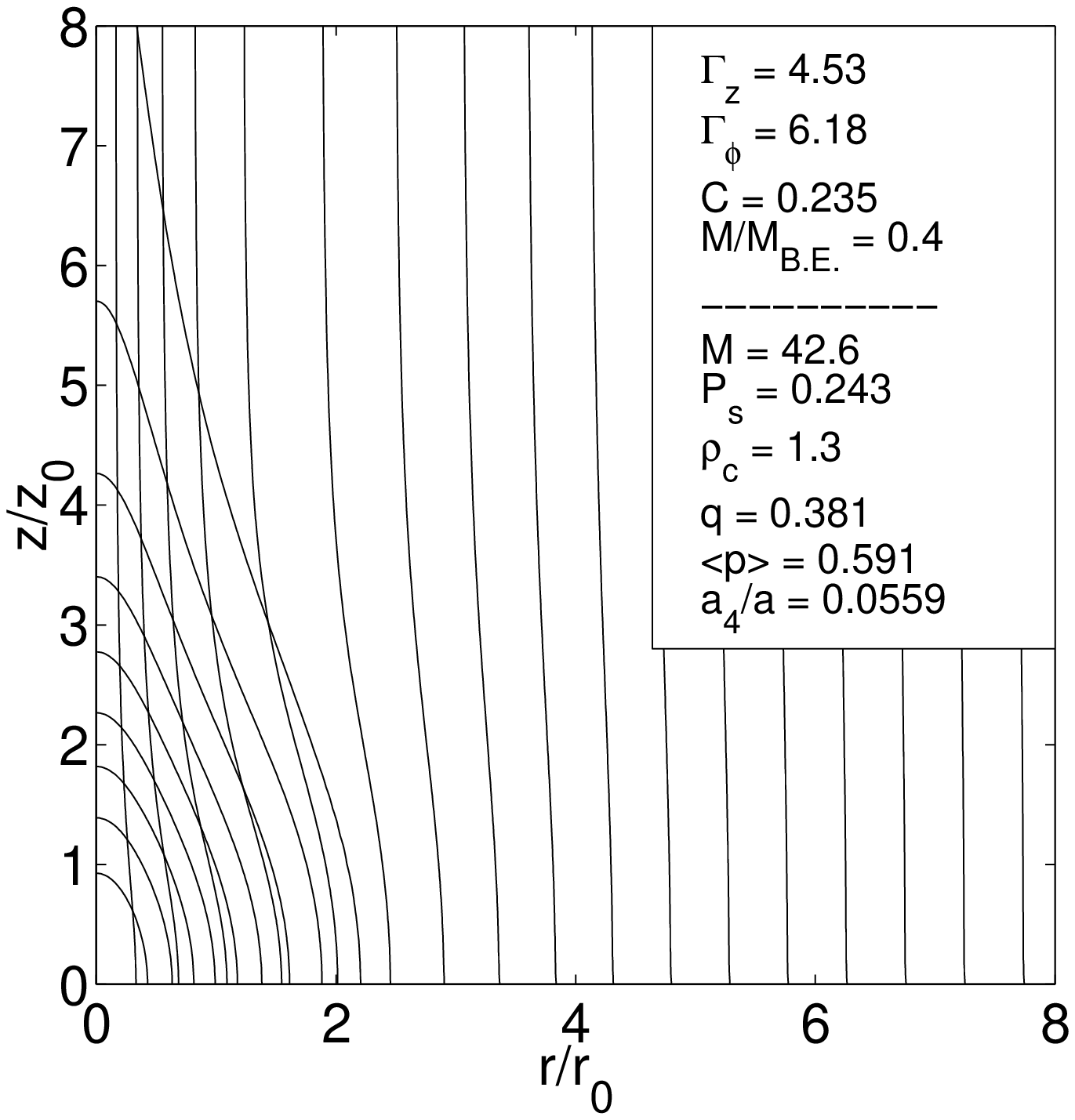,width=\textwidth}
\end{minipage}
\begin{minipage}{0.45\textwidth}
\psfig{file=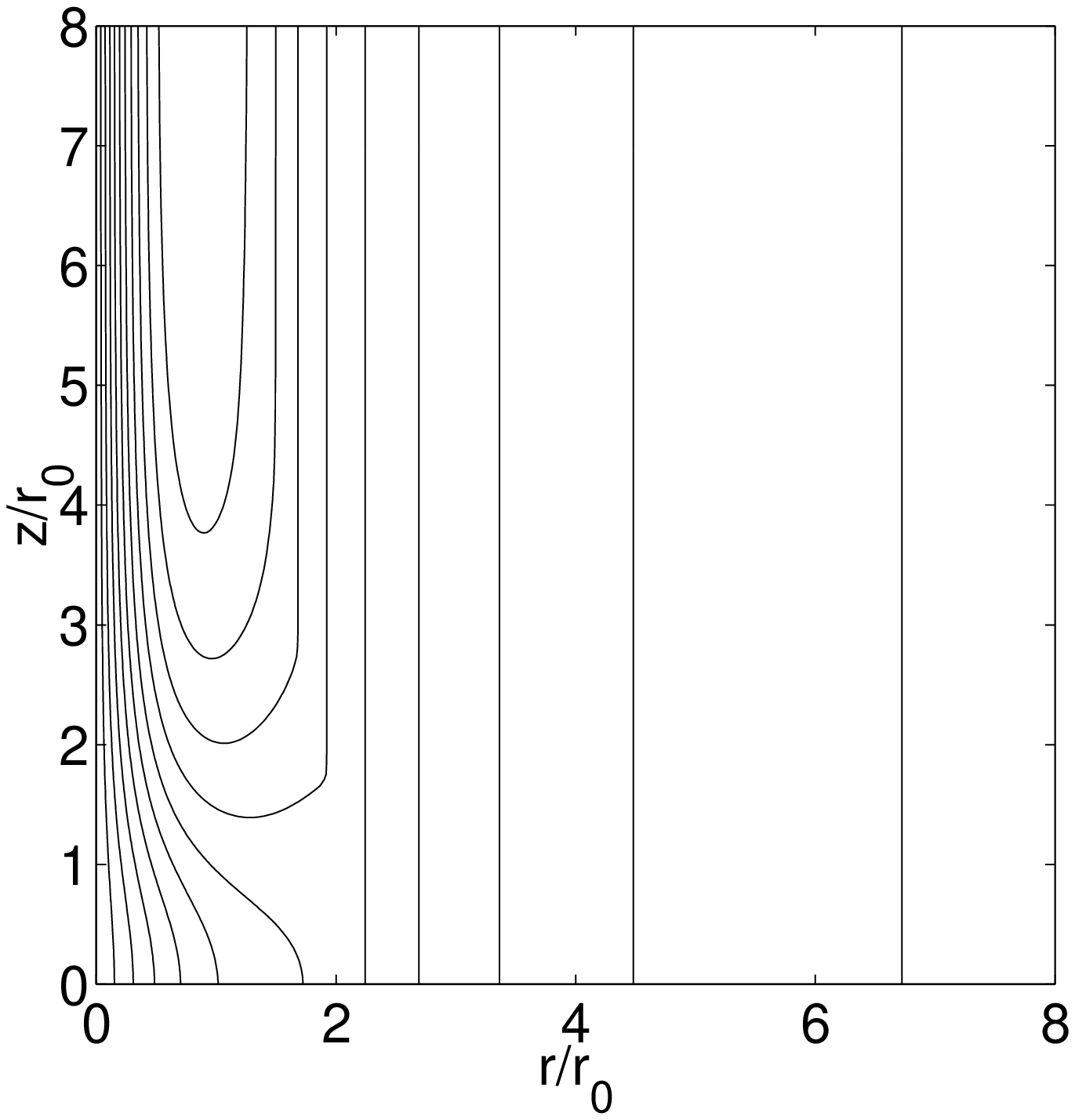,width=\textwidth}
\end{minipage}

\begin{minipage}{0.45\textwidth}
\psfig{file=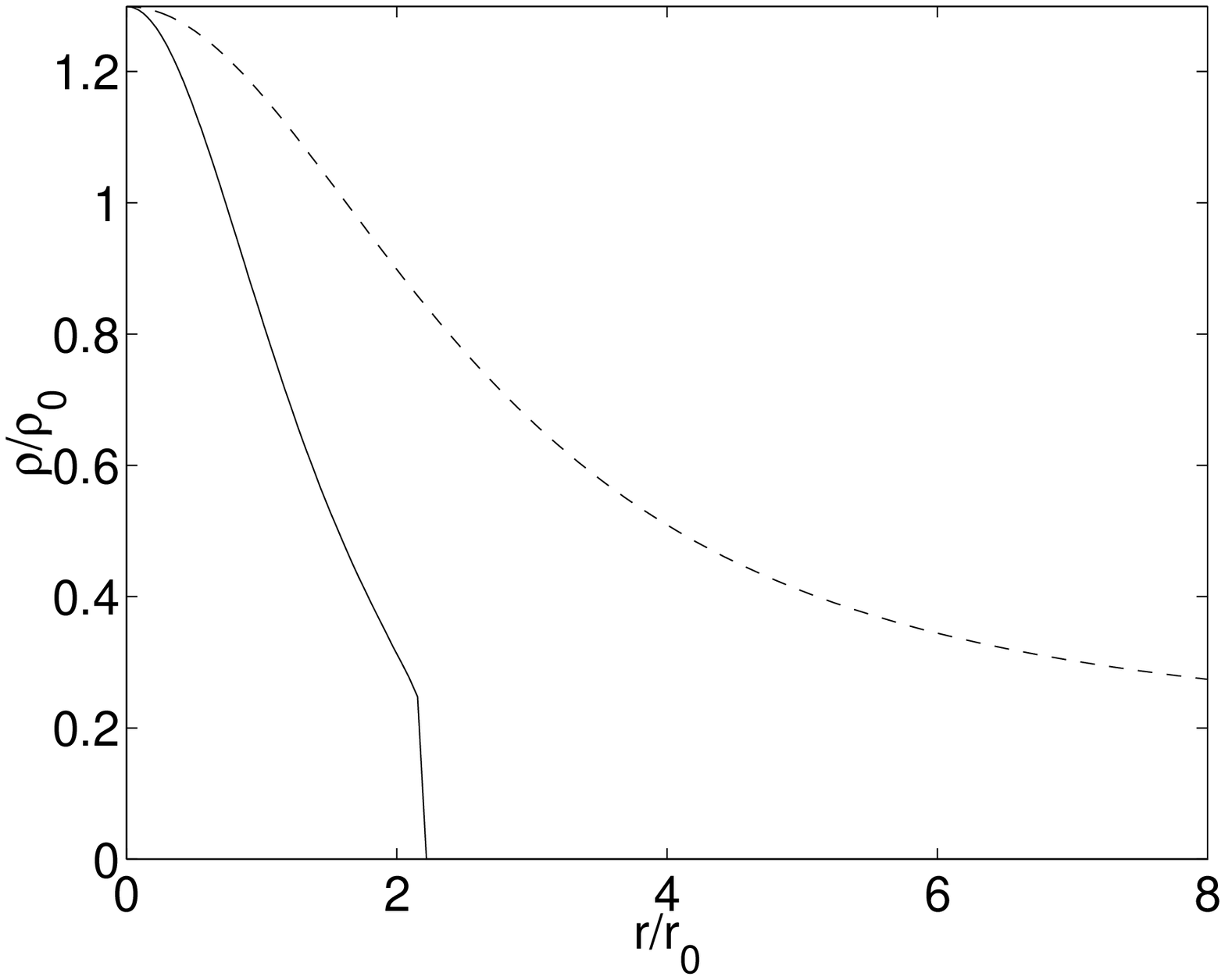,width=\textwidth}
\end{minipage}
\begin{minipage}{0.45\textwidth}
\psfig{file=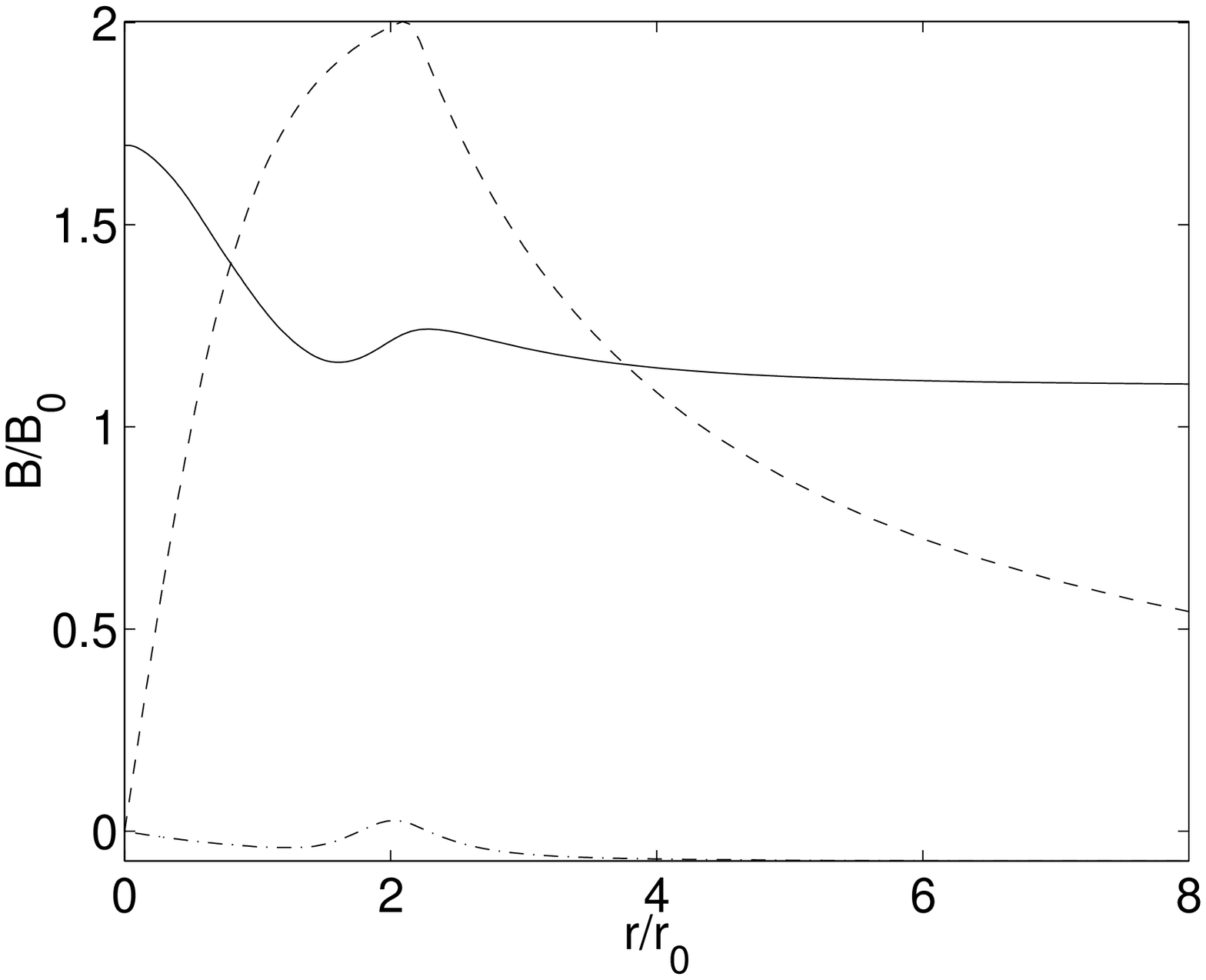,width=\textwidth}
\end{minipage}
\caption{Same as Figure \ref{fig:sample1}, but for a solution whose toroidal field component is stronger than the poloidal field
over much of the core.}
\label{fig:sample2}
\end{figure}

The model shown in Figure \ref{fig:sample2} contains regions where the toroidal field
becomes significantly stronger than the poloidal field.
The density contours are quite peaked towards the axis of symmetry, with $a_4/a>0.04$
for some models like the one shown.
While some models with this sort of behaviour are fairly elongated, it is interesting that most 
are actually less elongated than those with strong poloidal fields.  

\subsection{Trends in the Model Parameters}
\label{sec:trends}
\begin{figure}[ht]
\psfig{file=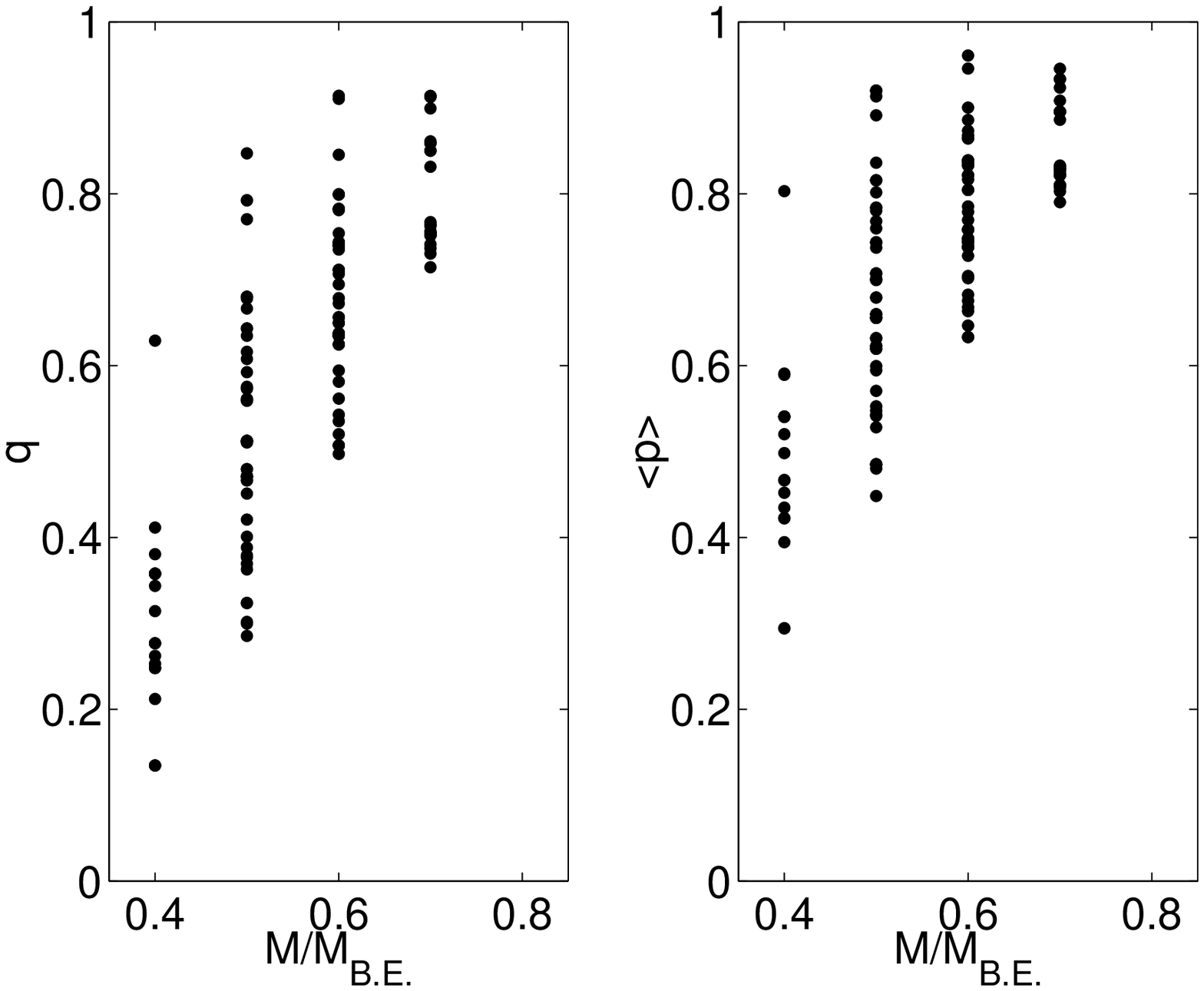,width=0.95\linewidth}
\caption{We show how the intrinsic and mean projected axis ratios depend on the ratio of the mass to the critical mass $\Mbe$ for
non-magnetic cores.  Gravitationally dominated cores with masses near $\Mbe$ are much less elongated than those with smaller
masses.  The error bars on $p$ are suppressed for clarity.}
\label{fig:MC4}
\end{figure}

\begin{figure}[ht]
\psfig{file=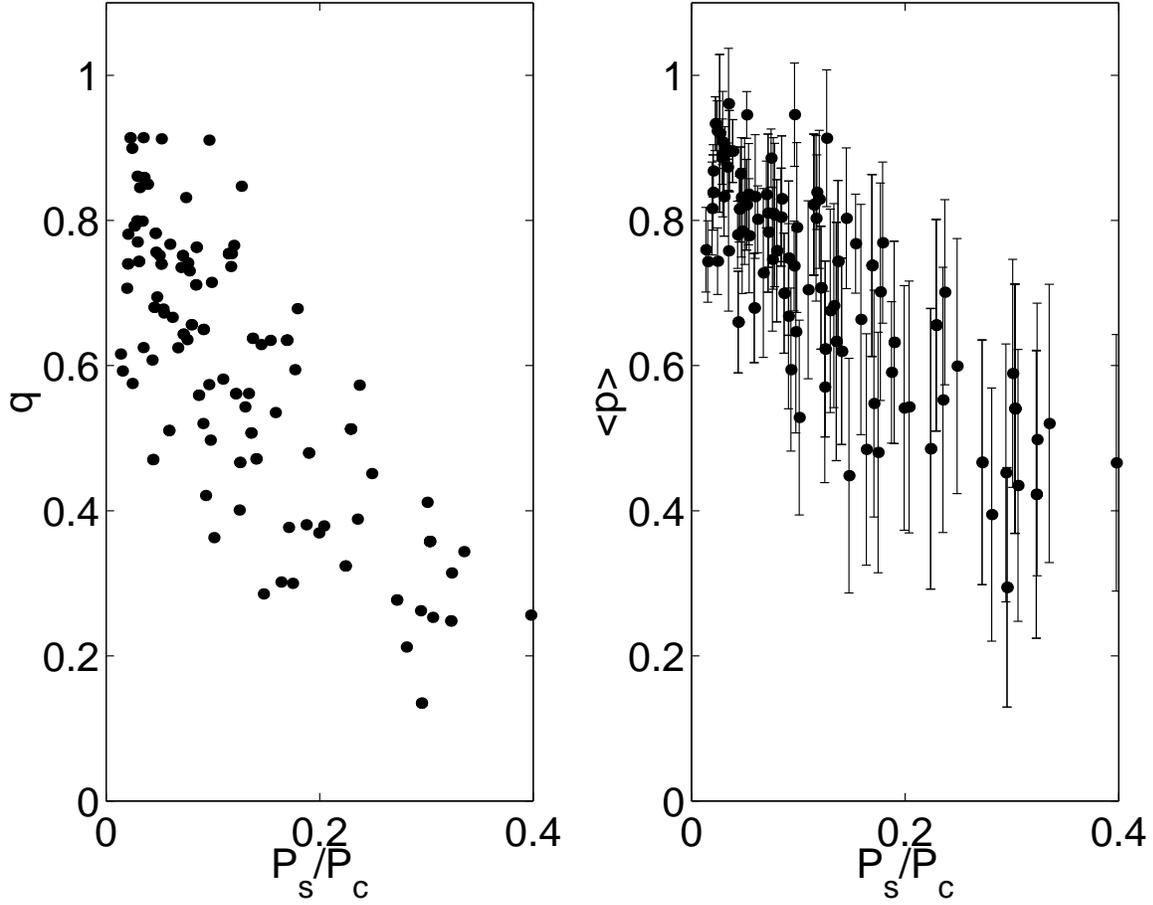,width=0.95\linewidth}
\caption{We show how the intrinsic and mean projected axis ratios depend on the ratio of the external pressure to the pressure at
the centre of the core.  Pressure dominated cores with relatively large values of $\Ps/\Pave$ are more elongated than less
pressure dominated models.}
\label{fig:MC5}
\end{figure}

\begin{figure}[ht]
\psfig{file=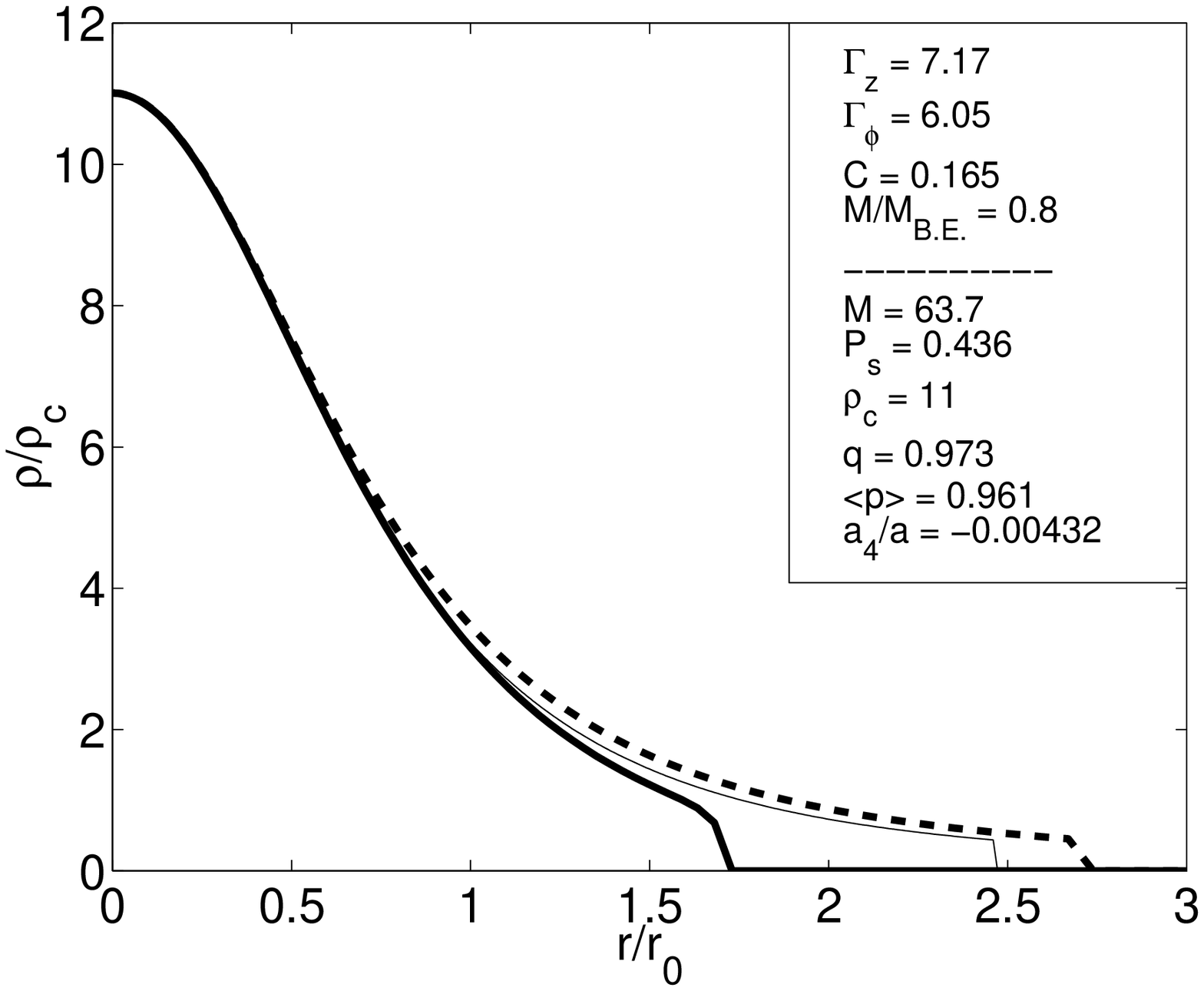,width=0.95\linewidth}
\caption{We show the most massive core obtained in our sample of 152 converged solutions.  The heavy solid line and dashed line
represent density cuts taken radially in the midplane and along the axis of symmetry respectively.  Both cuts agree well with
an isothermal sphere (thin solid line) in the inner, gravitationally dominated region, but deviate significantly in the pressure
dominated region near the surface.}
\label{fig:LE}
\end{figure}

Figures \ref{fig:MC4} and \ref{fig:MC5} illustrate that
our models generally become more elongated as the mass is decreased or as $\Ps/\Pave$ is increased.
The reasons for these trends are straightforward.
Given that rotation is negligible and the velocity dispersion is isotropic, gravitationally dominated cores should
be spherical.  We find nearly spherical, gravitationally dominated cores when the mass is large and the central
density is high.  The helical magnetic field can most effectively shape low mass cores, in which self-gravity is less important and the
surface pressure is a significant fraction of the central pressure.  The most massive core in our sample of converged solutions
is shown in Figure \ref{fig:LE}.  We show the radial density structure of a purely hydrostatic isothermal
sphere overlaid on radial and axial cuts of the density from our massive solution.  
We find that the inner, most gravitationally dominated regions
are described by an isothermal sphere to a high degree of accuracy
within $\sim 0.5~r_0$ of the centre (maximum deviation $=1.3\%$).
Beyond this radius, the magnetic field becomes important relative to gravity,
resulting in an axis ratio of 0.64 for the contour describing the surface of the core.

\begin{figure}[ht]
\psfig{file=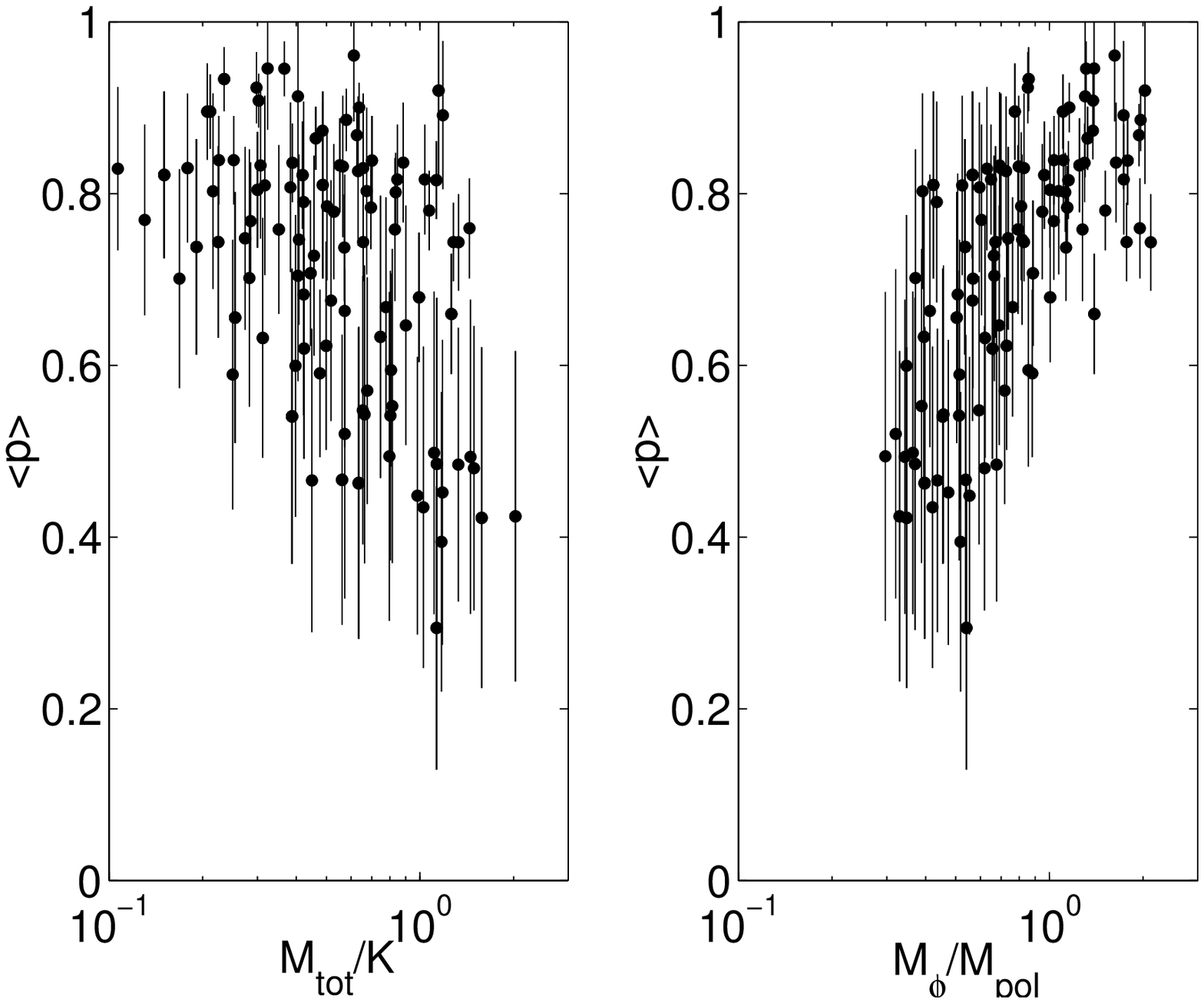,width=\linewidth}
\caption{ {\bf Panel (a)} (left): we show how the projected axis ratio $\pave$ depends on the
ratio of the magnetic to kinetic energy $M_{tot}/K$.  $M_{tot}$ does not exceed $\sim 2K$ for any of our 
models.  {\bf Panel (b)} (right): We show how $\pave$ depends on ratio of the toroidal to poloidal magnetic energy
$M_\phi/M_{pol}$.  Significantly elongated models have $M_\phi<M_{pol}$.}
\label{fig:MC6}
\end{figure}

Figure \ref{fig:MC6} verifies our assertion in Section \ref{sec:gallery} that only modest toroidal fields
are required to produce substantially elongated cores.
In panel (a), we show that the projected axis ratio $\pave$ decreases as the ratio $M_{tot}/K$ increases, where
$M_{tot}$ and $K$ are respectively the total magnetic energy
\footnote{
The magnetic energy $M_{tot}$ is not the full magnetic energy term from the virial theorem, since we do not include
any surface terms.  Our definition is more relevant to Zeeman measurements, since $M_{tot}$
is related to the total average field strength within the core.  We also
note that $M_{tot}/K$ is analogous to our parameter $X$ in FP1.
}
and the kinetic energy, defined by
\bea
M_{tot} &\equiv& \frac{1}{8\pi} \int_V B^2 ~dV \nn\\
K &\equiv& \frac{3}{2} \int_V P ~dV.
\label{eq:MK}
\eea
Models with $\pave < 0.6$ have $M_{tot}/K\appgeq 0.4$ in our model, but $M_{tot}/K \appleq 2$ even for our most elongated models.
Therefore, the field strengths required by our model are consistent with approximate equipartition between kinetic and magnetic energies,
which has often been inferred from observations (eg. Myers \& Goodman 1988a,b).
Panel (b) shows that most of the magnetic energy is in the poloidal component of the field for significantly 
elongated cores, with $\pave <0.6$.  We plot $\pave$ against $M_\phi/M_{pol}$, where $M_\phi$ and $M_{pol}$ are 
respectively the toroidal and poloidal contributions to the total magnetic energy in equation \ref{eq:MK}. 
As a general trend, we find that the most 
elongated models have the lowest values of $M_\phi/M_{pol}$, and that most models with $\pave <0.6$ have
$0.3\appleq M_\phi/M_{pol} \appleq 1$.  This trend might seem paradoxical because the toroidal field is responsible 
for the elongation.  However, models with energetically significant toroidal
fields must also have strong poloidal fields to suppress MHD instabilities.

\section{Sequences of Models with Varying Mass and External Pressure}
\label{sec:sequences}
Our emphasis in Sections \ref{sec:gallery} and \ref{sec:trends} was on exploring the full range of equilibrium states that are
accessible to our models.
In this section, we select a few filament models and examine sequences of core models that are obtained
by gradually increasing the mass and the external pressure.

The sequences shown in this section terminate
when the mass or pressure become sufficiently high that no equilibrium can be found due to a central density that increases
with each iteration of our numerical code, apparently without bound.  The maximum mass or pressure found in
this way clearly represents the
Bonnor-Ebert critical mass or pressure, which are both
reduced by the magnetic field.  These findings are consistent with the results of
Tomisaka (1991) and Habe et al (1991).

\subsection{Effect of Mass}
\label{sec:varymass}
\begin{figure}[ht]
\psfig{file=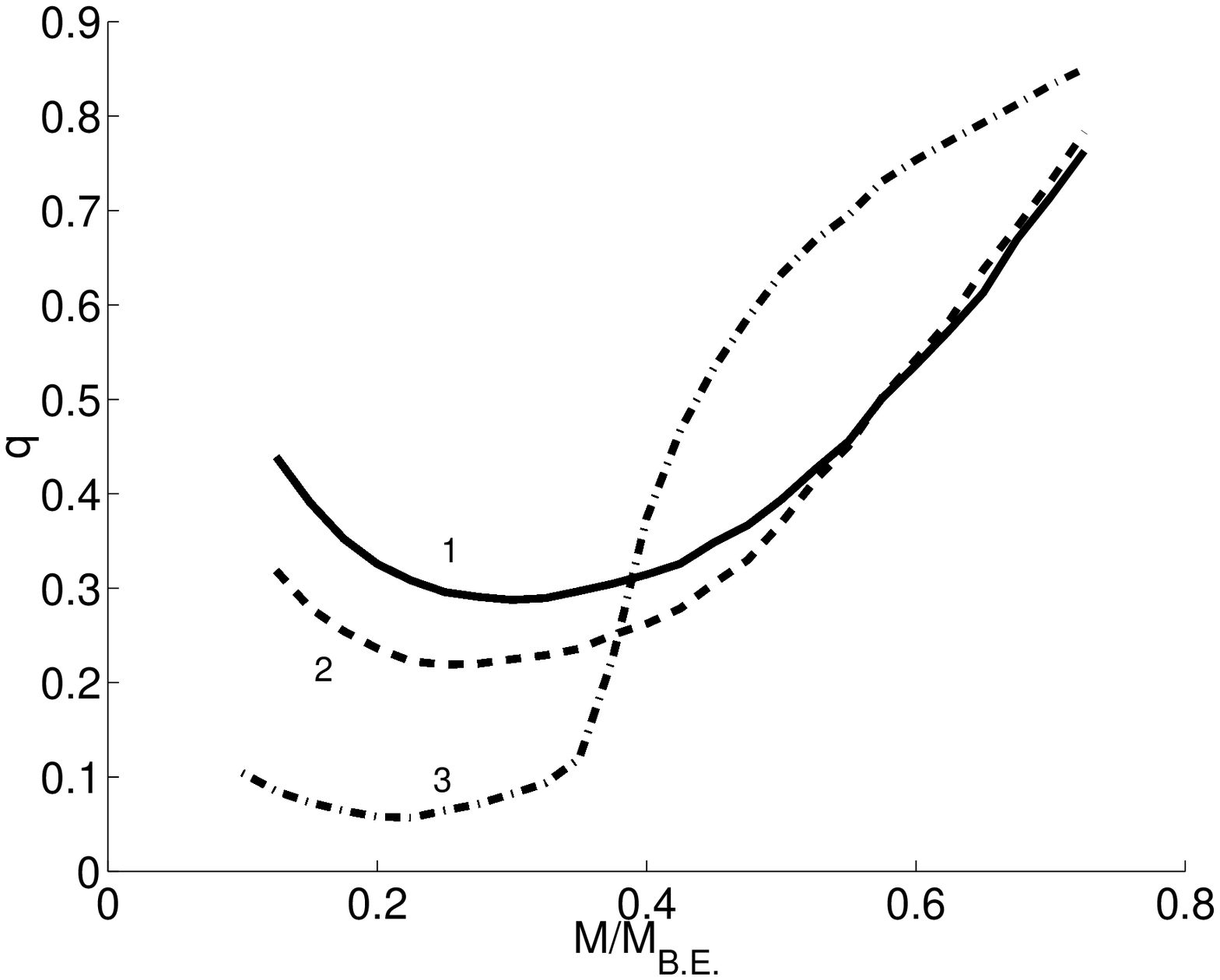,width=\linewidth}
\caption{We show the effect of varying the masses of cores on $q$ while holding all other parameters fixed.
The curves are labeled to correspond with the column labeled ``Sequence'' in Table \ref{tab:BSEQ}.}
\label{fig:BSEQ}
\end{figure}

Figure \ref{fig:BSEQ} shows how the intrinsic axis ratio $q$ varies with mass for
core models obtained from 3 different parent filaments.  We have normalized the mass by the Bonnor-Ebert critical mass $M_{B.E.}$,
to provide a convenient reference point for the mass.
The filament parameters $\Gz$, $\Gphi$, and $C$ are given
in Table \ref{tab:BSEQ}, as well as the external pressure $P_{ext}$ and $M_{B.E.}$.
We find that the axis
ratio $q$ decreases with increasing mass, when the mass is very low (less than a few tenths of $M_{B.E.}$).
These weakly self-gravitating equilibria are strongly pinched by the toroidal field
and more closely resemble short filaments than cores.
Once self-gravity becomes important (when the mass is a significant fraction of $M_{B.E.}$),
the axis ratio decreases with increasing mass.
Our models become nearly spherical when $M_{core} \appgeq 0.75 M_{B.E.}$.

The sequences shown in Figure \ref{fig:BSEQ} all terminate at masses that are about 20\% less than $\Mbe$,
although we have attempted to compute models with greater masses.  This is consistent with our finding that
very few models converge when $M_{crit}/\Mbe>0.8$ (see Figure \ref{fig:slices}).
Therefore, the critical mass $M_{crit}<\Mbe$ for our models, as a result of the pinch provided by the helical
magnetic field.

\begin{table}
\begin{tabular}{|c|c|c|c||c|c|}
\hline
Sequence	& $\Gz$	& $\Gphi$	& $C$	& $P_{ext}$	& $M_{B.E.}$ \\
\hline
1	& 7.17	& 6.05	& 0.165	& 0.436	& 79.6  \\
2       & 6.76  & 6.45  & 0.200 & 0.353 & 88.6  \\
3	& 4.53	& 6.18	& 0.235	& 0.243	& 106.6 \\
\hline
\end{tabular}
\caption{We give the dimensionless parameters used to compute the sequences of models shown in Figure \ref{fig:BSEQ}.
Parameters to the left of the double vertical line are input parameters, while the quantities to the right
are derived parameters of the model.}
\label{tab:BSEQ}
\end{table}

\subsection{Bonnor-Ebert Stability and the Effect of the External Pressure}
We have previously considered the external pressure bounding the core to be the same as that
bounding the parent filament.  However, for these sequences only, we vary  the external pressure
to determine how squeezing prolate cores affects their shapes.  This procedure also allows us to determine
the Bonnor-Ebert stability of our models.

\begin{figure}[ht]
\psfig{file=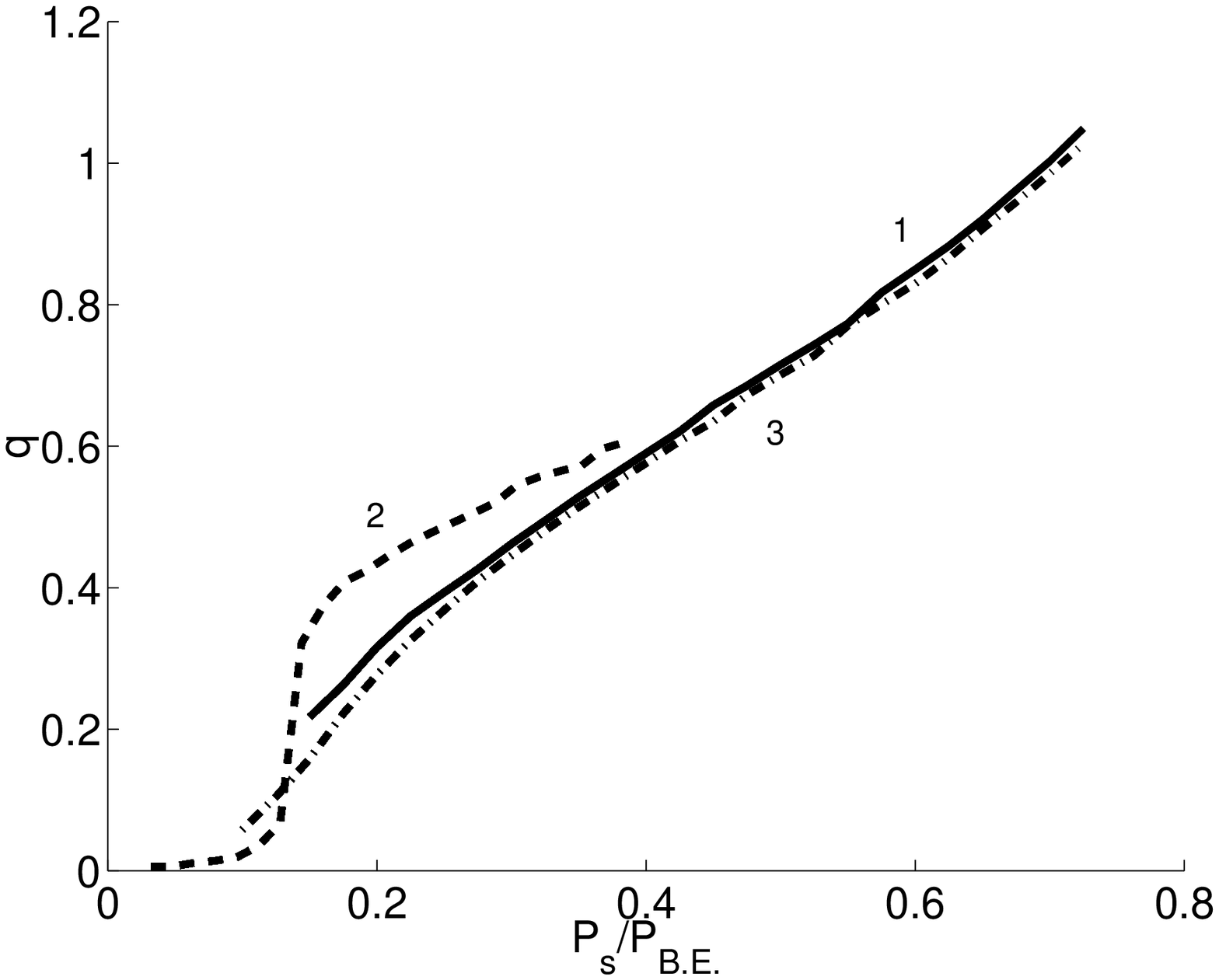,width=\linewidth}
\caption{We show the effect of varying the external pressure on $q$ for cores of constant mass.
The curves are labeled to correspond with the column labeled ``Sequence'' in Table \ref{tab:CSEQ}.}
\label{fig:CSEQ}
\end{figure}

Figure \ref{fig:CSEQ} shows how the shape of a core with fixed mass, $\Gz$, $\Gphi$, and $C$ changes as
the external pressure is varied.
We have held the mass fixed and varied the external pressure, which we have normalized by the
Bonnor-Ebert critical pressure $P_{B.E.}$, defined by
\be
P_{B.E.}=1.39\frac{\sigma^8}{G^3 M_{core}^2}
\ee
for convenience.
The parameters describing the parent filament are given in Table \ref{tab:CSEQ}, as well as the mass $M_{core}$ and $P_{B.E.}$.
We find that our models are most elongated when the surface pressure $\Ps$ is well below the Bonnor-Ebert critical pressure $P_{B.E.}$.
Cores become strongly self-gravitating, and therefore nearly spherical, when $\Ps$ is comparable to $P_{B.E.}$.

We find that all of the sequences terminate at 
pressures that are well below the Bonnor-Ebert critical pressure.  Therefore, the critical pressure $P_{crit}<\Pbe$, as a result
of the helical field squeezing our models in concert with the external pressure.

\begin{table}
\begin{tabular}{|c|c|c|c|c||c|}
\hline
Sequence	& $\Gz$	& $\Gphi$	& $C$	& $M_{core}$ & $P_{B.E.}$ \\
\hline
1	& 7.17	& 6.05	& 0.165	& 39.8	& 0.519 \\
2	& 4.53	& 6.18	& 0.235	& 42.6	& 0.496 \\
3	& 6.76	& 6.45	& 0.200	& 44.3	& 0.484 \\
\hline
\end{tabular}
\caption{We give the dimensionless parameters used to compute the sequences of models shown in Figure \ref{fig:CSEQ}.
Parameters to the left of the double vertical line are input parameters, while the quantity to the right
is a derived parameter of the model.}
\label{tab:CSEQ}
\end{table}

\section{A Singular Model for Prolate Cores}
\label{sec:SIE}
We have found a very simple singular isothermal solution for a prolate core containing a purely toroidal magnetic field.
While a purely toroidal field is admittedly not very realistic, the solution is interesting because it probably represents the simplest
possible model of a magnetized prolate core, and is in fact a generalization of the well-known singular isothermal sphere.
From a physical standpoint, this model is useful because it demonstrates that a relatively moderate toroidal field,
with plasma $\beta\ge 1$ everywhere, can result in elongated cores with axis ratios that are in good agreement
with the observations.

We assume a logarithmic gravitational potential, in cylindrical coordinates $(r,\phi,z)$, of the form
\be
\Psi = \sigma^2 \ln{ \frac{(r^2+\epsilon^2 z^2)}{2} } \label{eq:Psising} \\
\ee
where $\epsilon$ is the axis ratio of the isopotential surfaces, and $\sigma$ is assumed to be constant
\footnote{
See Binney and Tremaine (1994) equation 2-54a for a similar potential used in the context of galactic disks.
}.
This is the potential of a singular isothermal sphere when $\epsilon=1$, and describes singular prolate models when $\epsilon<1$.
In Appendix \ref{app:SIEderivation}, we show that the corresponding density and toroidal
magnetic field are given by 
\bea
\rho &=& \frac{\sigma^2}{2\pi G}\frac{\epsilon^2\left[r^2+(2-\epsilon^2)z^2\right]}{(r^2+\epsilon^2 z^2)^2} \label{eq:rhosing}\\
\Bphi &=& \frac{2 \sigma^2}{G^{1/2}} \frac{r \sqrt{1-\epsilon^2}}{(r^2+\epsilon^2 z^2)} \label{eq:Bphising},
\eea
where all quantities are the same as in our numerical models (See Section
\ref{sec:SCFM}).  It is clear that more elongated solutions (smaller values of $\epsilon$) 
are consistent with larger values of $\Bphi$.

We note that choosing $\epsilon=1$ recovers the well-known singular isothermal sphere solution,
for which $\Bphi$ vanishes.  It is interesting that the density vanishes when $\epsilon=0$.  Therefore, the
solution does not reduce to a singular model for filaments in this limit (See also FP1).
An important feature of equation \ref{eq:Bphising} is that   
there is no solution when $\epsilon>1$.  Hence, there are no oblate solutions that are 
consistent with our singular model.

\begin{figure}[ht]
\begin{minipage}{.49\linewidth}
\psfig{file=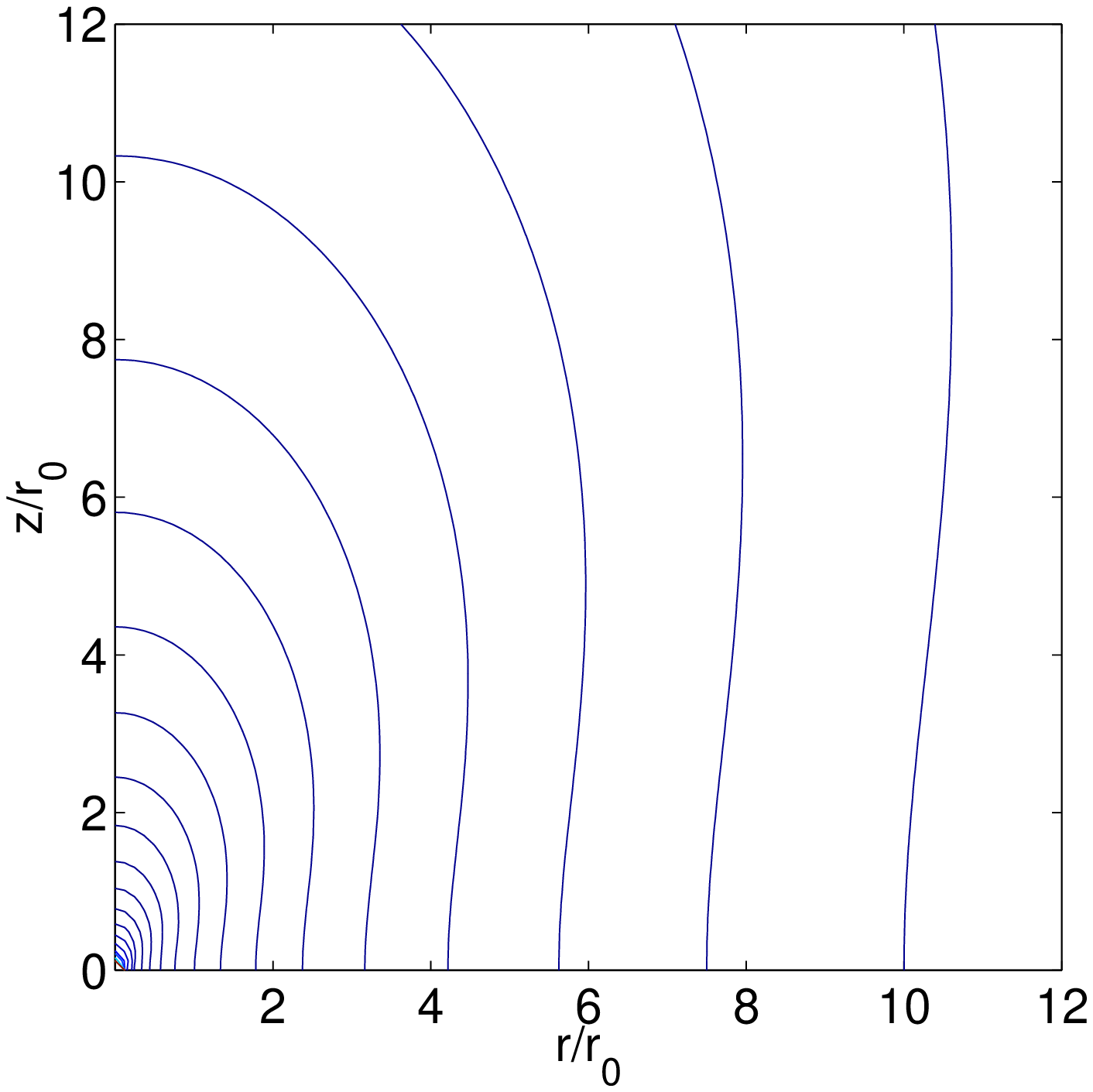,width=\textwidth}
\end{minipage}
\begin{minipage}{.49\linewidth}
\psfig{file=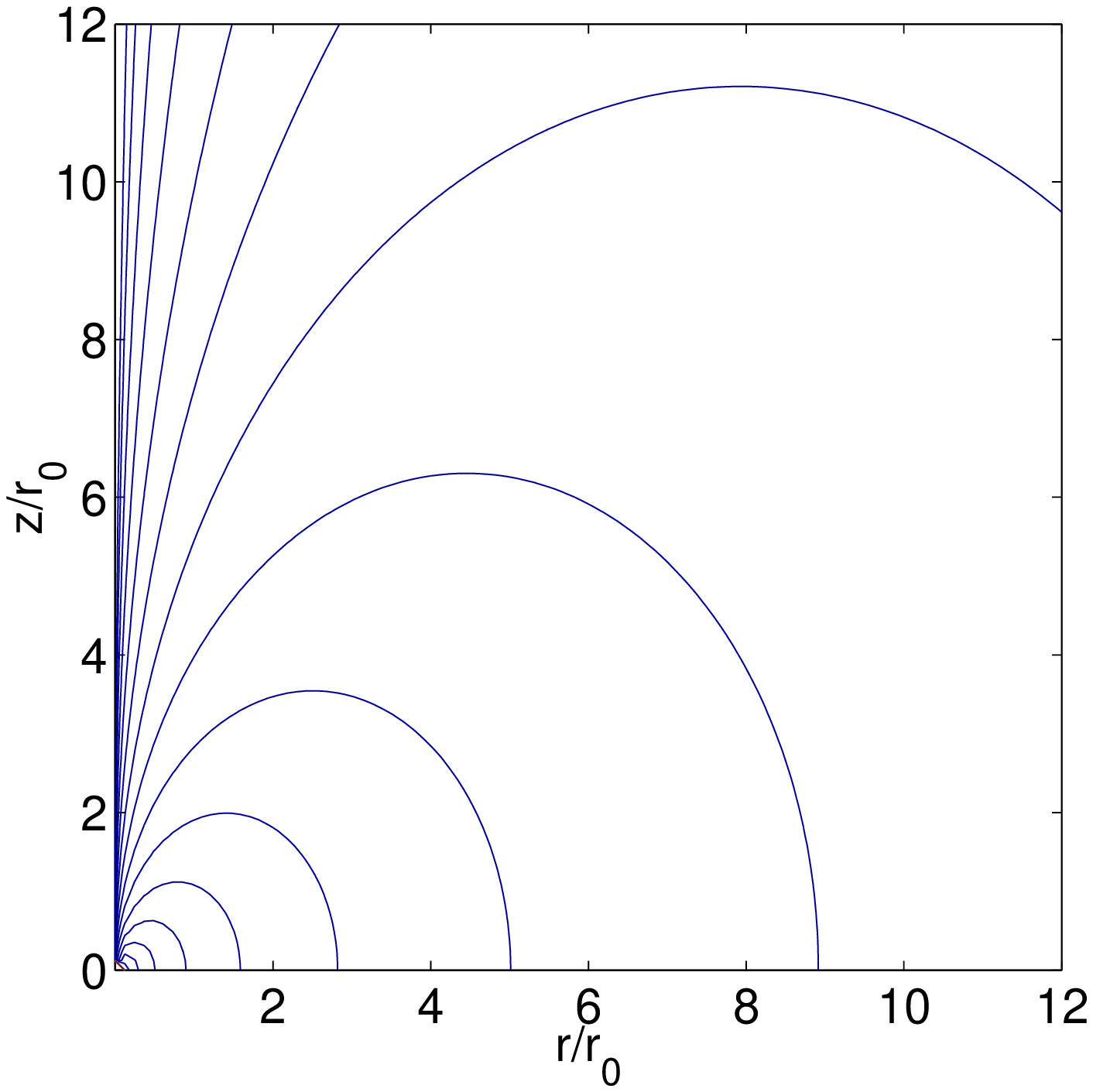,width=\textwidth}
\end{minipage}
\caption{We show density contours (top, panel a) and toroidal field intensity contour (bottom, panel b) for the
singular model for prolate cores with a purely toroidal field.}
\label{fig:sing}
\end{figure}

The plasma $\beta$ parameter is given by
\be
\beta=\frac{\epsilon^2}{1-\epsilon^2} \left[1+\frac{(2-\epsilon^2)z^2}{r^2}\right],
\ee
which has a maximum of $\epsilon^2/(1-\epsilon^2)$ in the midplane, where $z=0$.
Therefore, $\epsilon=1/\sqrt{2}$ results in a moderately magnetized model with $\beta\ge 1$ everywhere,
which we show in Figure \ref{fig:sing}.
We find that the density contours of this singular model have an axis ratio of $\approx 0.46$
(See Section \ref{sec:MC} and Appendix \ref{app:proj} for a discussion of our method of estimating axis ratios.).  
This result is in good agreement
with the mean intrinsic axis ratios that are typically derived for samples of cores (See Ryden 1996).

A simple numerical example illustrates that relatively moderate toroidal fields can result in significant elongation.
We consider the magnetic field at a distance of $0.1~pc$ from the centre of the $\epsilon=1/\sqrt{2}$ 
model discussed above.  This distance corresponds to the radius at which the thermal and non-thermal velocity dispersions
are approximately equal for low mass cores (Fuller \& Myers 1992, Casseli \& Myers 1996).  Therefore,
$\sigma\approx 0.26~km~s^{-1}$, assuming a temperature of $10~K$ for the gas.  
Equation \ref{eq:Bphising} requires a maximum strength of only $\approx 12~\mu G$ for $\Bphi$ at this distance.  

The toroidal field in the singular model
is strongest in the midplane, which squeezes the gas so that the constant density surfaces
are slightly pinched in.  This is markedly different from our non-singular numerical solutions,
which have their maximum toroidal field off the midplane.  However, models with purely toroidal fields, 
such as our singular solution, are a separate class of solutions that cannot be
obtained from the self-consistent field equations (discussed in Section \ref{sec:SCFM}), even 
in the limit of a vanishing poloidal field.  The reason for this difference is that
azimuthal component of the magnetohydrostatic equilibrium equation (equation \ref{eq:magnetostatic})
is automatically satisfied for purely toroidal field geometries.  
In the more general case of a helical field, torsional equilibrium requires $\bp$ to be constant along poloidal field lines,
so that the class of allowed solutions is somewhat more restricted.

\section{Discussion and Conclusions}
\label{sec:discussion}
We have previously (FP1) suggested that some filamentary molecular clouds may contain helical
magnetic fields, and that such filaments are unstable against fragmentation along their lengths (FP2).
Some fragments may dynamically evolve into equilibrium states that can be identified as cloud cores.
The central point of the present work is to demonstrate that most of the possible equilibrium states that arise from our
filament models are prolate, and that many have realistic axis ratios.  

It is very encouraging that we are able to explain several of the observed features of filamentary clouds and their
cores within a self-consistent theoretical framework.
Future tests of our models may be provided by measurements
of polarized dust grain emission.
A useful test of our models will be to predict the polarization due to both our filament
models (FP1) and the prolate core models derived in this paper.
This is done in a forthcoming study.  Perhaps even more importantly, the central assumption of our model, which
posits a constant flux to mass loading, requires direct confirmation.  This could one day be feasible
with sufficiently high spatial resolution Zeeman measurements.

Previous observations of the shapes of cores have concentrated mainly on their axis ratios (Myers et al. 1991, Ryden 1996).
However, we have found that the shapes of cores can usefully be described by a second 
observable parameter $a_4/a$, which quantifies
the deviation of the density contours from ellipses.  
The toroidal field in our model results in $a_4/a>0$, which is consistent with surface density
contours that are somewhat football-shaped (ie. peaked along the major axis).  
Therefore a detailed isophotal analysis of cores may provide important
clues to their internal magnetic structures.

Core models that exceed the critical mass for stability in the sense of Bonnor (1956) and Ebert (1955)
are doomed to collapse.  The dynamical collapse of one of our prolate core models from
a critical initial state is a difficult problem that should be addressed by future numerical simulations.

\section{Summary}
\label{sec:summary}
\noi 1.  Most of the core models that are consistent with our previous models of filamentary molecular clouds
are found to be prolate.  A wide range of axis ratios is possible, with $0.1\appleq q\appleq 0.9$ and
$0.3\appleq \pave\appleq 1$. \\
2. Both density and surface density profiles are found to be slightly football-shaped 
(ie. peaked toward the symmetry axis) in our models,
as quantified by the $a_4/a$ parameter of Bender \& M\"ollenhoff (1987).\\
3. The Bonnor-Ebert critical mass is reduced by about 20\% in our models by the toroidal field.\\
4. We find that models are generally more elongated for low values of the mass ratio $M/\Mbe$ and 
the pressure ratio $\Ps/\Pbe$. \\
5. Only modest toroidal fields are required to produce elongated cores.  
Most of our models have $\Bphi<\Bz$ nearly everywhere. 

\section{Acknowledgements}
J.D.F. acknowledges the financial support of McMaster University during this research.
The research of R.E.P. is supported by a grant from the Natural Sciences and Engineering Research
Council of Canada.

\appendix
\section{The Mean Projected Axis Ratio and $a_4/a$ Shape Parameter}
\label{app:proj}
For each of the models discussed in the previous sections, we have calculated the surface density at various
angles of inclination.  By assuming that cores are oriented randomly, we calculate a distribution
function for the axis ratio and the $a_4/a$ shape parameter
of the half maximum surface density contour, from which we obtain their mean values and standard deviations.  
We work in Cartesian coordinates $(x',y',z')$, with the origin at the centre of the
core, the $y'$ axis along the line of sight
to a distant observer, and the $x'-z'$ plane parallel to the plane of the sky.  We define a second set of Cartesian coordinates
$(x,y,z)$ fixed relative to the core, with $z$ corresponding to the axis of symmetry.
The $z$ axis is assumed to be tilted by an inclination angle $i$ relative to the plane of the sky and towards the observer.
The coordinates are then related by
\bea
x' &=& x \nn\\
y' &=& y\cos{i}-z\sin{i} \nn\\
z' &=& y\sin{i}+z\cos{i}.
\eea
Since the cylindrical radius (relative to the symmetry axis) is given simply by $r=\sqrt{x^2+y^2}$, we interpolate
$\rho(r,z)$ over the primed coordinates to obtain $\rho(x',y',z')$.  
The surface density $\Sigma(x',z')$ is then obtained by integrating over $y'$.  
We then compute the contour of the half maximum surface density, from which we
estimate the axis ratio by fitting an ellipse as described in Section \ref{sec:MC}.
We also compute the $a_4/a$ shape parameter, using the method discussed in
Bender and M\"ollenhoff (1987).
We repeat this procedure at many different inclination angles to determine the projected axis
ratio $p$ and $a_4/a$ as a function of $i$.

The $z$ axis is assumed to be oriented in a random direction $(\theta,\phi)$ relative to the $z'$ axis.
The inclination angle is related to $\theta$ and $\phi$ by
\be
\cos{i}=1-\sin^2\theta \sin^2\phi.
\label{eq:i}
\ee
The probability $dp_\phi$ of the $z$ axis being oriented between $\phi$ and $\phi+d\phi$ is
\be
dp_\phi=\frac{d\phi}{2\pi}.
\ee
Integrating, the probability of the $z$ axis being oriented between $0$ and $\phi$ is given by $\phi/2\pi$,
so that
\be
\phi=2\pi\xi
\ee
if $\xi$ is a uniformly distributed random variable.
Similarly, it is easy to show that
\be
\cos{\theta}=1-2\xi
\ee
for another random variable $\xi$.  We generate a large number of realizations (typically $10^5$) of the
orientation $(\theta,\phi)$, from which we calculate the orientation $i$ by equation \ref{eq:i}.  We
interpolate to find $p$ and $a_4/a$ for each realization, and calculate the mean values and standard
deviations of the resulting distributions.

\section{Derivation of the Singular Isothermal Model for Prolate Cores}
\label{app:SIEderivation}
Working in the dimensionless variables introduced in equation \ref{eq:dimensionless} and assuming that the
magnetic field is purely toroidal, the equation of
magnetohydrostatic equilibrium (equation \ref{eq:magnetostatic}) can be written as
\be
8\pi r^2 e^{-\Psi}\nabla Q=\nabla\bp^2
\label{eq:magnetostatic2}
\ee
where $\bp$ is related to $\Bphi$ by equation \ref{eq:bphi}.
The variable $Q$ has the same form as in equation \ref{eq:hydro2}, but is no longer contsrained to be a function of $\Phi$:
\be
Q=\rho e^\Psi.
\label{eq:Qdef2}
\ee
We note that the $r$ and $z$ components of equation \ref{eq:magnetostatic2} can be combined into the form of
a Jacobian determinant, which implies that $\bp$ can be considered as a function of $Q$ alone.

The potential for the singular isothermal sphere model, in our dimensionless variables, is given
by $\Psi=\ln(R^2/2)$, where $R$ is the spherical radius.  Therefore, we consider distorted logarithmic potentials
of the form
\be
\Psi=\ln\left(\frac{r^2+\epsilon^2 z^2}{2}\right)
\label{eq:Psi2}
\ee
where $\epsilon$ is the axis ratio of the isopotential surfaces.
The density is easily obtained from Poisson's equation (equation \ref{eq:poisson2}):
\be
\rho=2\epsilon^2\frac{\left[r^2+(2-\epsilon^2)z^2\right]}{(r^2+\epsilon^2 z^2)^2},
\label{eq:rho2}
\ee
which appears in dimensional form as equation \ref{eq:rhosing}.

Inserting equations \ref{eq:Psi2} and \ref{eq:rho2} into equation \ref{eq:Qdef2}, we find that $Q$ can be written
in the form 
\be
Q(\zeta)=\frac{\epsilon^2\left[ 1+(2-\epsilon^2)\zeta^2\right]}{1+\epsilon^2\zeta^2},
\label{eq:Qxi}
\ee
where 
\be
\zeta\equiv z/r.
\label{eq:xi}
\ee
Since $\bp$ is a function of $Q$ alone, we can consider both $Q$ and $\bp$ to be functions
of $\zeta$ in equation \ref{eq:magnetostatic2}.  This equation becomes an ordinary differential equation in for $\bp(\zeta)$:
\be
\frac{d\bp}{d\zeta}=-\frac{8\pi}{1+\epsilon^2\zeta^2}\frac{dQ}{d\zeta}.
\ee
Substituting $Q(\zeta)$ into this equation and integrating, we find that
\be
\bp^2=\frac{16\pi(1-\epsilon^2)}{(1+\epsilon^2\zeta^2)^2}.
\ee
The toroidal field $\Bphi$ is directly obtained from this equation by replacing both $\zeta$ and $\bp$ with their  
respective definitions (equations \ref{eq:xi} and \ref{eq:bphi}).  The final dimensional form is written out in 
equation \ref{eq:Bphising}.

\end{document}